\documentclass[desactivate]{aa}

\usepackage{natbib,twoopt}
\usepackage{multirow}
\usepackage[breaklinks=true]{hyperref}
\usepackage{amsmath}
\usepackage{amssymb}	
\usepackage[T1]{fontenc}
\usepackage{booktabs}
\usepackage{threeparttable}

\usepackage{tabularx}
\usepackage{graphicx}
\usepackage{txfonts}

\bibpunct{(}{)}{;}{a}{}{,} 
\usepackage{siunitx}
\begin{document}

\title{Confirmation of the presence of a CRSF in the  NICER spectrum of X 1822-371}


   \author{R. Iaria\inst{1}, T. Di Salvo\inst{1}, A. Anitra\inst{1}, 
   C. Miceli\inst{1,2}, F. Barra\inst{1}, W. Leone\inst{3,1}, L. Burderi\inst{4}, A. Sanna\inst{4}  and  A. Riggio\inst{4}}

 \institute{Dipartimento di Fisica e Chimica - Emilio Segrè,
 Universit\`a di Palermo, via Archirafi 36 - 90123 Palermo, Italy
 \and
             INAF/IASF Palermo, via Ugo La Malfa 153, I-90146 Palermo, Italy
             \and
 Department of Physics, University of Trento, Via Sommarive 14, 38122 Povo (TN), Italy 
  \and
         Dipartimento di Fisica, Universit\`a degli Studi di Cagliari, SP
Monserrato-Sestu, KM 0.7, Monserrato, 09042 Italy 
  }

  \abstract
   {}
   {X 1822-371 is an eclipsing  binary system with a period close to 5.57 hr and an orbital period derivative $\dot{P}_{\rm orb}$ of 1.42(3)$\times 10^{-10}$ s s$^{-1}$. The extremely high value of its $\dot{P}_{\rm orb}$ is compatible with a super-Eddington mass transfer rate from the companion star and, consequently, an intrinsic luminosity at the Eddington limit. The source is also an X-ray pulsar, it shows a spin frequency  of 1.69 Hz and  is in a spin-up phase with a spin frequency derivative of  $7.4 \times 10^{-12}$ Hz s$^{-1}$. Assuming a luminosity at the Eddington limit,  a neutron star magnetic field strength of  $B = 8 \times 10^{10}$ G is estimated. However, a direct measure of $B$ could be obtained observing a CRSF in the energy spectrum.
    Analysis of \textit{XMM-Newton} data suggested the presence of a cyclotron line at 0.73 keV, with an estimated magnetic field strength of $B=(8.8 \pm 0.3) \times 10^{10}$ G.  
   }
   {Here we analyze the 0.3-50 keV broadband spectrum of  X 1822-371 combining a 0.3-10 keV  NICER spectrum and a 4.5-50 keV \textit{NuSTAR}  spectrum   to investigate the presence of a cyclotron absorption line  and the complex continuum emission spectrum.}
   {The  NICER  spectrum confirms the presence of a cyclotron line  at 0.66 keV.  The continuum emission is modeled with a Comptonized component,  a thermal component associated with the presence of an accretion disk truncated at the magnetospheric radius of 105 km and a reflection component from the disk blurred by relativistic effects.  
  }
  {We confirm the presence of a cyclotron line at 0.66 keV inferring a NS magnetic field of $B = (7.9\pm 0.5) \times 10^{10}$ G  and suggesting that the Comptonized component originates in the accretion columns.  }

  \authorrunning{R. Iaria et al.}

  \titlerunning{Confirmation of a CRSF at 0.7 keV in the spectrum of X 1822-371}
  
  

   \maketitle
%

\section{Introduction}
The low-mass X-ray binary system (LMXB) X 1822-371 (4U 1822-37) is a persistent eclipsing source with an orbital period of 5.57 hr, hosting an accreting X-ray pulsar with a spin frequency close to 1.69 Hz \citep{jonker_01} and in a spin-up phase with  a derivative of $\dot{\nu}=(7.39 \pm 0.03) \times 10^{-12}$ Hz s$^{-1}$ \citep{Mazzola_19}.  X 1822-371 belongs to the class of accretion disc corona (ADC) sources \citep{white_82}, with an inclination angle between \ang{81} and \ang{84} \citep{heinz_01}. 
The distance to this source was estimated to be between 2-2.5 kpc by \cite{mason_82} using infrared and optical observations. The 0.1-100 keV unabsorbed luminosity is $1.2 \times 10^{36}$ erg s$^{-1}$, adopting a distance of 2.5 kpc \citep{iaria_01_a}.
{Recently, using {\it Gaia} data, \cite{Arnason21} estimated a  distance to X 1822-371 of $6.1^{+1.6}_{-2.7}$ kpc, which implies that the observed luminosity should be a factor by six larger than that estimated in the literature.}

The most recent orbital ephemeris of the source X 1822-371 was reported by \cite{Anitra_21}, who suggested that the orbital period derivative is $\dot{P}_{\rm orb}=(1.42 \pm 0.03) \times 10^{-10}$ s s$^{-1}$ adopting a quadratic ephemeris.
This high orbital period derivative cannot be explained by a conservative mass transfer at the accretion rate inferred from the observed source luminosity.  A highly 
non-conservative mass transfer is needed  at a rate up to seven times the Eddington limit for a neutron star  (NS) mass of 1.4 M$_\odot$ \citep[see][]{burderi_10,bay_10,iaria_11,Mazzola_19}. In this scenario, the intrinsic luminosity produced by the source is likely at the Eddington limit (a few $10^{38}$ erg s$^{-1}$), two orders of magnitude higher than the observed luminosity of $10^{36}$ erg s$^{-1}$.

 Assuming the neutron star (NS) is in spin equilibrium, \cite{jonker_01}  showed that  the NS magnetic field   is close to $B \simeq 8 \times 10^{10}$ G for an intrinsic luminosity of $10^{38}$ erg s$^{-1}$ while  $B \simeq  2  \times 10^{16}$ G for an intrinsic luminosity of $10^{36}$ erg s$^{-1}$. Finally, they suggested a $B$ value close to $10^{12}$ G, discussing the spectral models reported in the literature up to that time, which yields an intrinsic luminosity of the source of $10^{37}$ erg s$^{-1}$  given the measured $ \dot{P}_{pulse}/P_{pulse}$  value.

However, a direct measure of the magnetic field strength $B$   is possible only by observing a cyclotron resonance scattering feature (CRSF) in the spectrum. \cite{sasano_14}, studying the \textit{Suzaku} broadband spectrum of X 1822-371, suggested the presence of a CRSF at 33 keV and a corresponding value of $B \simeq 3 \times 10^{12}$ G.   \cite{iaria_15}  showed that the 
B-value proposed by \cite{sasano_14} is not consistent with the spin-up  regime  of the source. 
By studying the \textit{XMM} spectrum of X 1822-371,  \cite{iaria_15} identified a  possible CRSF at 0.73 keV  corresponding to a value of $B = (8.8 \pm 0.3) \times 10^{10}$ G.  
The authors suggested that the observed Comptonized component could be produced in the accretion column onto the NS magnetic caps, and  estimated a magnetospheric radius of $\sim 250$ km.

\cite{Iaria_13}  suggested a geometry of the source consisting of an optically thick corona close to the neutron star identified by a Comptonization component in the source spectrum and an optically thin extended corona with an optical depth of $\tau = 0.01$. The optically thin corona was introduced for three reasons: 1)  since the intrinsic luminosity is at  the Eddington limit while the observed one is a hundred times weaker, an optically thin extended corona with  $\tau =0.01$ would scatter along the line of sight one percent of the intrinsic luminosity (the source is observed with an edge-on geometry); 2)  the optically thin corona must be extended because the observed eclipses are partial; 3) the NS spin  pulsation is observed. If the corona were extended and optically thick, the pulsation would be diluted by photon scatterings in the corona while assuming the existence of  an optically thin corona with  $\tau = 0.01$   the  travel time of the photons through the optically thin corona ($\simeq 2$ s) is comparable to the spin  period ($\simeq 0.59$ s)  and part  of the pulsed photons can reach the observer after scattering. 

\cite{Bak17} proposed that the seed-photons spectrum (composed of a blackbody component plus a  power law with a cut-off energy at 10 keV and photon index $\Gamma=1$) is  scattered by an extended optically thin  corona with an optical depth ranging from 0.01 to 3.  According to the authors, their model  eliminates the optically thick Comptonizing region and  explains the system only with an optically thin (or moderately thick;  $\tau \simeq 0.01 - 3$) corona surrounding the X-ray pulsar.

Recently, \cite{Anitra_21} showed that the broadband X-ray spectrum of the source can be fitted with a model composed of: a thermal component associated with an 
accretion disk emission, a Comptonized component and a reflection component from the accretion disk. The emission from the central region of the system is obscured from direct view by the outer edge of the disk. However, it is scattered along the line of sight by an optically thin corona with optical depth of 0.01.

In this work, we analyze the broadband spectrum of X 1822-371 combining  a  NICER  spectrum   with an exposure time of 11 ks  and \textit{NuSTAR} spectrum  with an exposure time of 96 ks.  
We confirm  the presence of a cyclotron absorption line  close to  0.7 keV by  taking advantage of the  large effective area of    NICER   between 0.5 and 2 keV. 
 We confirm that the model discussed by \cite{Anitra_21}   fits the broadband spectrum well
 and suggest that the Comptonizing corona is placed at the accretion column over the magnetic caps of X 1822-371.

 \begin{figure}[!htbp]
   \centering
    \includegraphics[scale=.65]{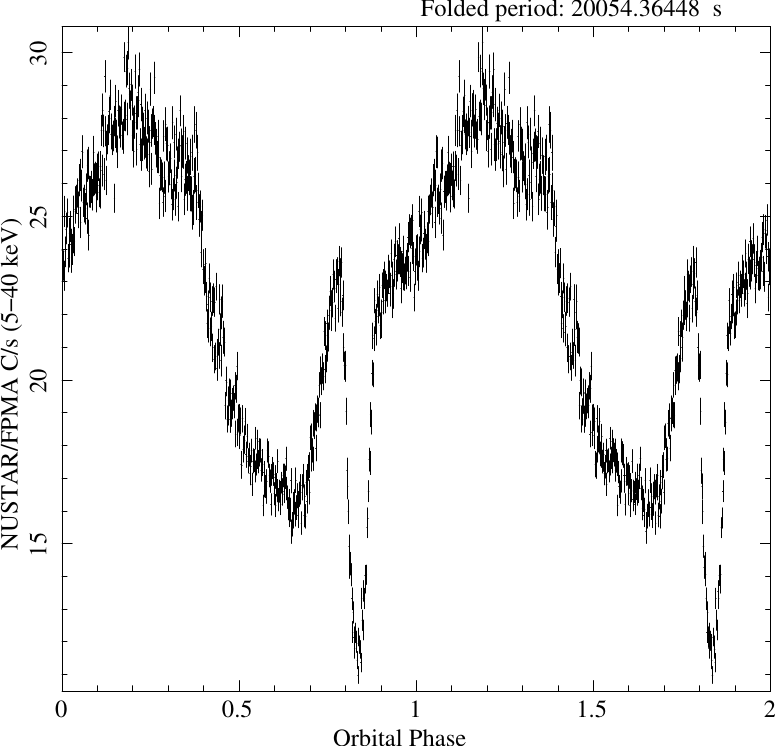}
    \caption{{\it NuSTAR}/FPMA folded light curve assuming as epoch $T_{\rm fold} = 59625$ MJD and an orbital period of  0.2321107 days. The period is divided into 512 bins. For clarity, two orbital phase cycles are shown. }
   \label{fig:folded}
\end{figure}

\section{Observations and data analysis}

 Neutron Star Interior Composition Explorer \citep[ NICER,][]{Gendreau_16} observed X 1822-371 on  2022 Jun 04  for a total exposure time of 11 ks (obsid. 5202780101). 
The primary instrument of  NICER  is the X-ray Timing Instrument (XTI), which is an array of 56 photon detectors that operate in the 0.2–12 keV energy range. We reduced the  NICER   data using the pipeline tool {\tt nicerl2} in NICERDAS v10 available with HEASOFT v6.31 and adopting the standard   filters; the used  calibration database version was  {\tt xti20221001}. 
{
During the observation the Focal Plane Modules (FPM) 63 and 43 were noisy; the tool {\tt nicerl2} rejected all the events collected by FPM 63 and generated GTIs for FPM 43 after data screening. In order to create the spectrum, we utilized the {\tt nicerl3-spect} tool to exclude FPM 14 and 34 due to elevated detector noise.}
 The 3-10 keV light curve was extracted using the tool {\tt nicerl3-lc}.
The source spectrum and the {\tt scorpeon} background file  were  extracted using the tool  {\tt nicerl3-spect} and setting the options {\tt bkgformat=file}. We added the systematic error suggested by the  NICER  calibration team   to the spectrum\footnote{\url{https://heasarc.gsfc.nasa.gov/docs/nicer/analysis_threads/cal-recommend/}}, and grouped the data  using the ftool {\tt ftgrouppha}  applying an optimal rebinning {\tt grouptype=optmin} to have at least 25 counts per energy bin  \citep[{\tt groupscale=25},][]{Kaastra_16}.   

{\it NuSTAR} observed X 1822-371  from 2022 February 15  at 08:21:09, to February 17   at 14:26:19 (ObsID 30701025002), with a duration of 200 ks and a total exposure time of 96 ks. To  produce  the light curve, the source and the background  spectra  of FPMA and FPMB, we utilized the {\tt Nupipeline} and {\tt Nuproducts} scripts in HEASOFT. A circular region with a radius of 100 arcsec was chosen to extract  the events from the source.  The background events were extracted from a circular region within the Field of View (FOV) that is devoid of source photons. This circular region was selected from the same detector where the source is located  and has a radius of 100 arcsec.    After checking that the FPMA and FPMB spectra are similar in the 4-50 keV energy band we summed the two spectra using the ftool  \texttt{addspec}\footnote{question 27 in \url{https://heasarc.gsfc.nasa.gov/docs/nustar/nustar_faq.html}}, the summed spectrum  was  rebinned as done for the   NICER  spectrum.

\subsection{Update of the orbital ephemeris}
We produced the  {\it NuSTAR} light curve extracting the   FPMA events in the 5-40 keV energy range  and applying the  barycentric corrections to the event  arrival times using the ftool {\tt barycorr}. 
  \begin{figure}[!htbp]
   \centering
    \includegraphics[scale=.75]{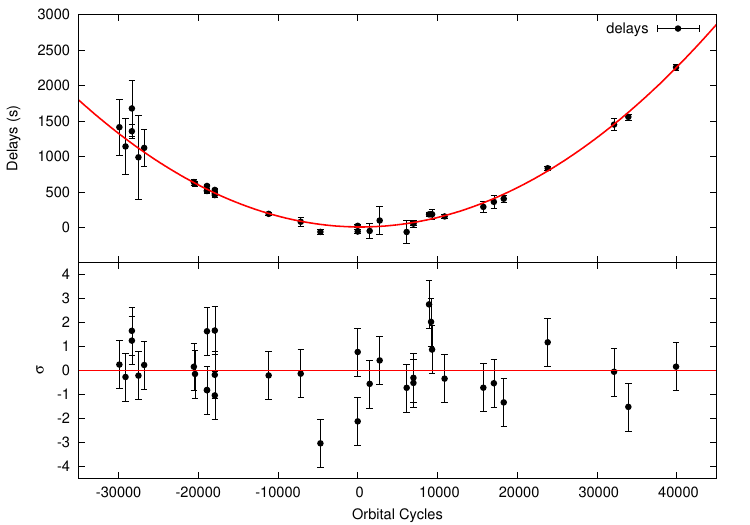}
    \caption{Delays vs. orbital cycles for the quadratic ephemeris (top panel).   We used the eclipse arrival times shown by \cite{Mazzola_19} (see tab. 1 in their work), the one obtained by \cite{Anitra_21} and that shown in the text. Residuals in units of $\sigma$ (bottom panel).}
   \label{fig:delay_tidal}
\end{figure}
  We folded the  light curve, adopting
a reference epoch T$_{\rm fold}$=59625  MJD and a reference
period of P$_{\rm fold}$=0.2321107 days (see  Fig. \ref{fig:folded}). To estimate the eclipse arrival time T$_{\rm ecl}$ we used the procedure proposed by \cite{burderi_10} finding T$_{\rm ecl}$=59625.19445(48) MJD/TDB.
Adopting a reference orbital period of P$_0$ = 0.232109571 days and a reference eclipse time T$_0$=50353.08728 MJD, we found that the number of orbital cycles $N$  is 39947   and the delay associated with the eclipse arrival time is 2259(41) s.  
To update the orbital ephemeris, we included the   eclipse arrival time shown above with those reported by \cite{Mazzola_19} and \cite{Anitra_21}.   
The delays as a function of orbital cycles reveal that the source exhibits stable expansion over 44 years of data (see top panel of \autoref{fig:delay_tidal}).

We fitted the delays as a function of cycles adopting the quadratic model $y=a+bN+cN^2$,   where $a=\Delta T_0$ is the correction to the adopted $T_0$, $b=\Delta P$ is the correction to the adopted $P_0$ and $c=1/2P_{0} \dot{P}$ allows us to estimate the orbital period derivative  $\dot{P}$. We  obtained a $\chi^2$(d.o.f.) of 47(32) \citep[similarly to what obtained by][]{Mazzola_19}.
  The best-fit quadratic curve  (red) and the corresponding residuals in units of $\sigma$ are shown, respectively, in the top and bottom panels of \autoref{fig:delay_tidal}.

    The updated orbital ephemeris are:
\begin{equation}
\label{eq:ephemeris}
\begin{split}
        T_{\rm ecl}= 50353.08740(13)  \; {\rm MJD/TDB} + 0.232109559(6)   N +\\ 
       + 1.655(30)\times 10^{-11} N^2,
\end{split}
\end{equation}
where the first and the second term represent the refined values of the reference epoch $T_{\rm 0,orb}$ and orbital period $P_{\rm 0,orb}$, respectively. The third term allows us to estimate an orbital period derivative of $\dot{P}_{\rm orb}= 1.426(26) \times 10^{-10}$ s s$^{-1}$. 
  
We added a cubic term in order to test the presence of a second derivative of the orbital period. By fitting the delays with a cubic function we found a $\chi^2(d.o.f.)$ of 43(31),  that is  the addition of a cubic term is not statistically significant 
({\bf $\sigma \sim 1.3$}).
 
Finally, we  verified whether a gravitational quadrupole coupling produced by tidal dissipation \citep{apple_94} could be detectable in our data as done by \cite{Mazzola_19}. To this aim, we added  a sinusoidal term to the quadratic model (hereafter LQS model). By fitting the delays with the LQS model  we obtained a  $\chi^2(d.o.f.)$ of 34(29), indicating that the addition of the sinusoidal term is not  statistically significant with a detection   of   2$\sigma$. 
 
\begin{figure}[!htbp]
   \centering
    \includegraphics[scale=.7]{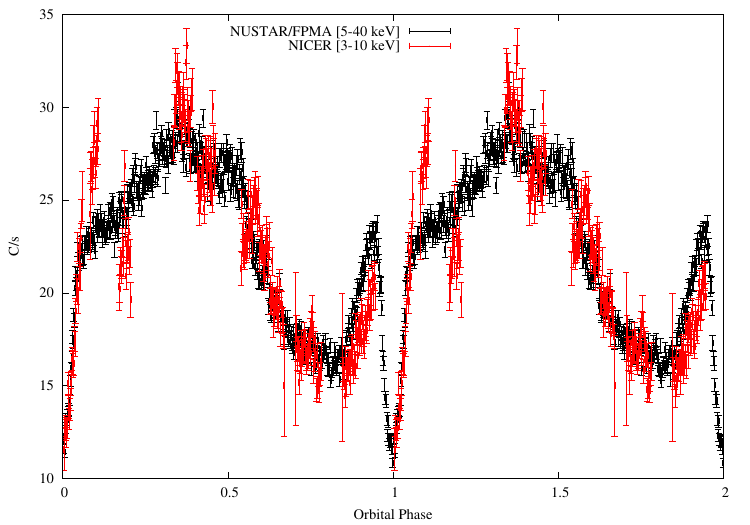}
    \caption{ 5-40 keV {\it NuSTAR}/FPMA  folded light curve (black curve) and 3-10 keV  NICER  folded light curve (red curve).   The folded light curves have 512 bins per peiod, each bin corresponds to 39.17 seconds.  }
   \label{fig:comparison_folded}
\end{figure}
    \begin{figure*}[!htbp]
   \centering
    \includegraphics[scale=.65]{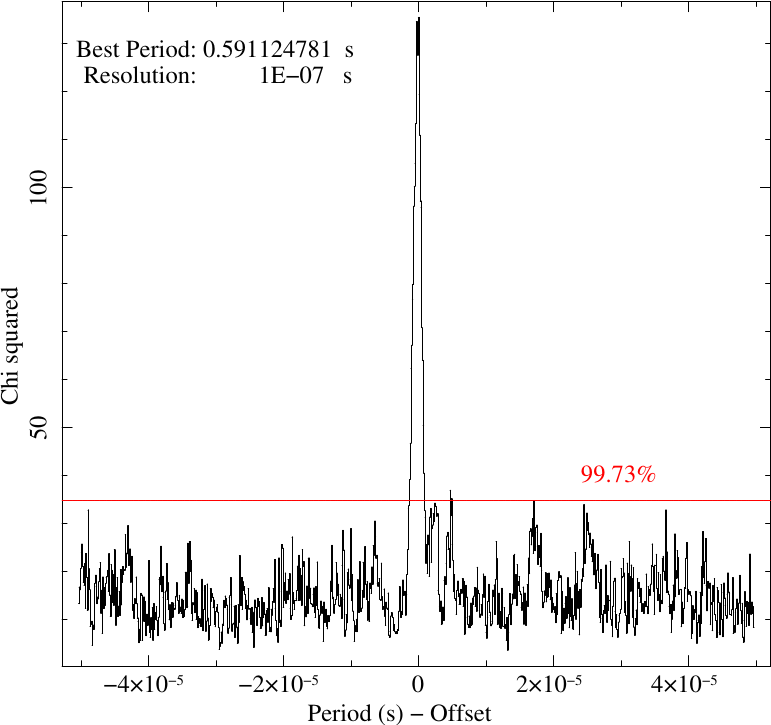}
        \includegraphics[scale=.65]{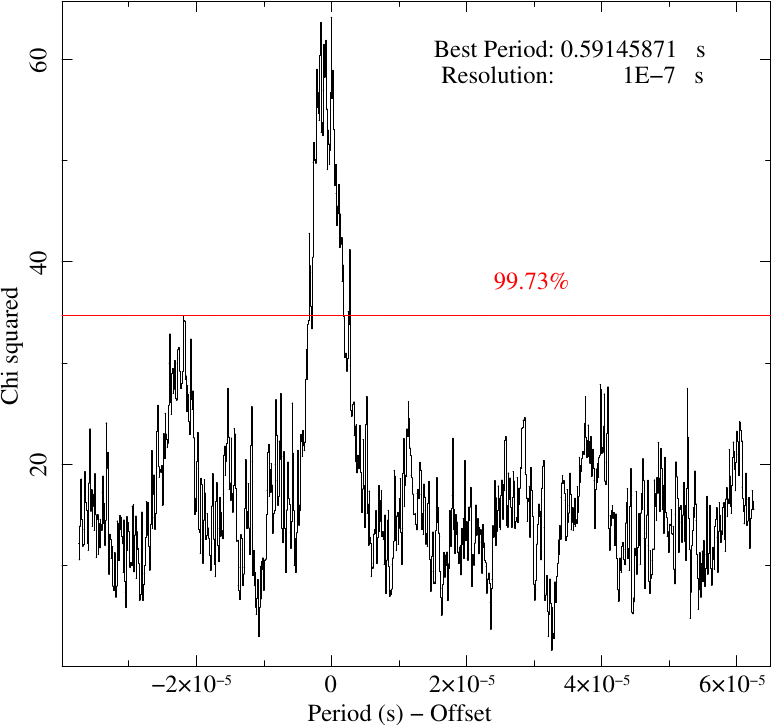}
    \caption{Folding search for periodicity in the 5–40 keV  {\it NuSTAR}/FPMA  light curve of the obsid. 30701025002 (left panel) and obsid. 30301009002 (right panel). The horizontal dashed line indicates the $\chi^2$   value of 34.72 at which we have the 99.73\% confidence level for a single trial, corresponding to a significance of 3$\sigma$.}
   \label{fig:efsearch}
\end{figure*}

The folded  {\it NuSTAR}/FPMA light curve,   imposing as null phase the passage at the eclipse and adopting the orbital period at the times of the observation, is shown in fig. \ref{fig:comparison_folded} (black curve). 

We applied the barycentric   correction to the 3-10 keV   NICER  light curve using the ftool {\tt barycorr} and folded it  using the eclipse arrival time (T$_e =59734.51859$ MJD/TDB) and the orbital period predicted by the ephemeris shown in  eq. \ref{eq:ephemeris} (P$_{orb} = 20054.36448$ s).
We show the folded light curve in fig. \ref{fig:comparison_folded}  (red curve), the orbital modulation
of the two light curves is similar along the orbital period. 
 \subsection{Spin-period search}
  \begin{table}[!htbp]
        \centering
        \caption{Log of the  spin periods} 
        \scriptsize
        \begin{tabular}{ llc }
      \toprule
        Time  (MJD) & P$_{spin}$ (s)& Reference \\  
      \midrule
50352.9(6) &  0.59325(2) & (1) \\
50993.2(6) &  0.59308949(2) & (2) \\
51018.9(7) &  0.59308615(9)  & (2) \\ 
51975.8(3) &     0.5928850(6)    & (3)\\
52032.1(6)      &     0.5928742(3)  & (2) \\    
52093(2)        &     0.59286187(5) & (4) \\
52435.5(2)      &     0.59279212(5) &  (2)\\
52496(8)        &     0.59278016(13) & (4)\\ 
52519.3(2)      &     0.59277244(1)   & (2) \\
52547.4(3)      &     0.59276601(4)  & (2) \\
52608.2(1)      &     0.59281075(3)   & (2) \\
52889(7)        &     0.59268544(15) & (4) \\ 
54010.5(5)      &    0.5924337(10)  & (5) \\  
55882(1)      &       0.59199982(6)  &  (2)\\
55892(4)      &       0.59199464(3) &  (2)\\ 
57818.4(4)    &     0.5915669(4) & (6) \\ 
58234(1)     &    0.59145871(8)  & (7) \\     
59626.4(1.1)    &    0.591124781(13)  & (7) \\   
    \bottomrule
\end{tabular}
    \begin{tablenotes}
\item[] References: (1) \cite{jonker_01},  (2) \cite{Bak17}, (3) \cite{iaria_15}, (4) \cite{Chou_16}, (5) \cite{sasano_14},  (6) \cite{Mazzola_19}, (7) this work. The times were calculated as  the mean value between the start and stop time of the observations. 
    \end{tablenotes}
        \label{tab:pspin}
    \end{table}
\begin{figure}[!htbp]
   \centering
    \includegraphics[scale=.65]{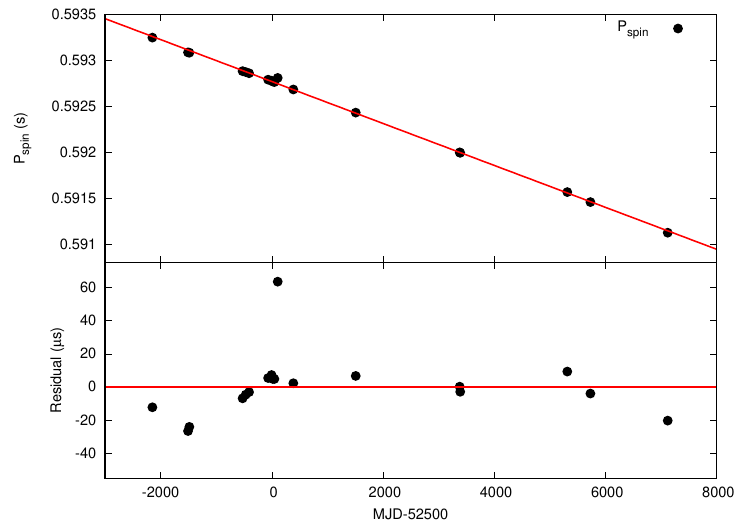}
    \caption{Long-term spin period change   (top panel) and   corresponding residuals in units of $\mu$s.   The linear decrease of the pulse period over time suggests that the NS is in a spin-up regime. }
   \label{fig:pspin}
\end{figure}

\begin{figure*}[!htbp]
   \centering
    \includegraphics[scale=.65]{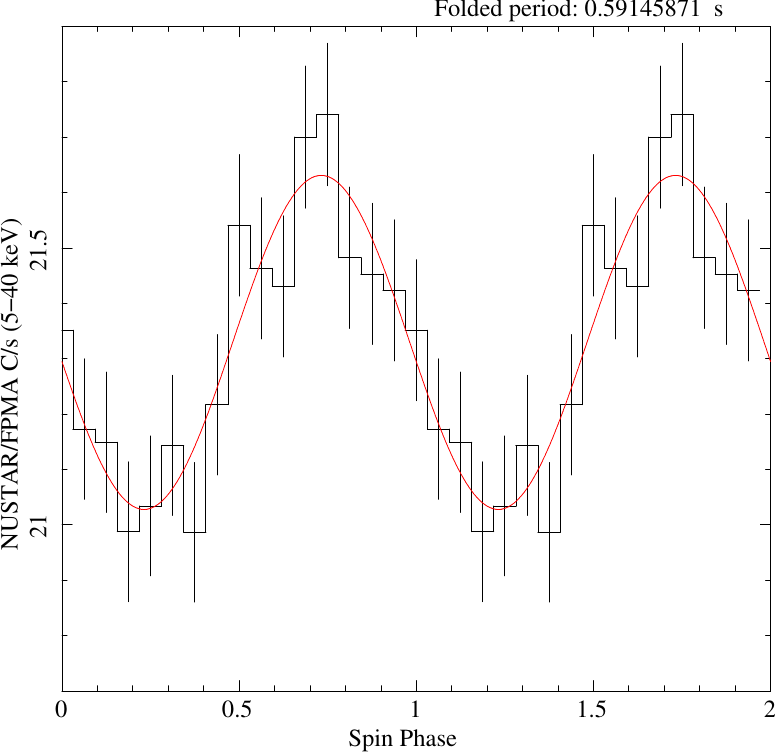}
        \includegraphics[scale=.65]{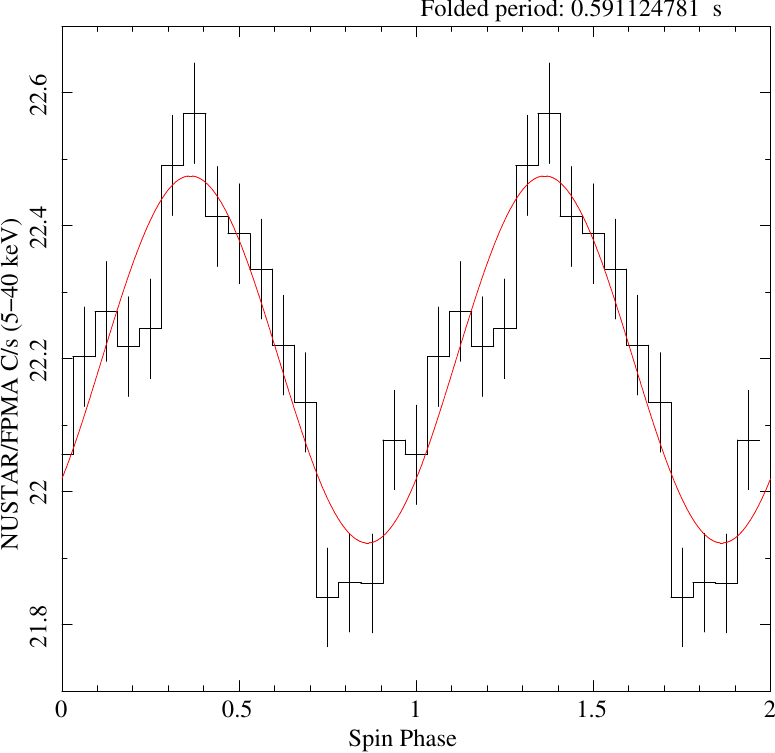}
    \caption{5-40 keV {\it NuSTAR}/FPMA  folded light curve  of the obsid. 30701025002 (left panel) and obsid. 30301009002 (right panel). The curves are folded to have 16 bins per period. Two periods are shown fro clarity. The red curves represent the best-fit model using a sinusoidal function.}
   \label{fig:profile}
\end{figure*}

We searched for the NS spin frequency in the  5-40 keV {\it NuSTAR}/FPMA light curve   binned at 0.01 s.    We corrected the event arrival times  for the binary orbital motion using $a \sin i = 1.006(5) $ lt-s \citep{jonker_01}, 
the eclipse arrival time shown above and the orbital period at the time of the observation. 

 We used  the ftool {\tt efsearch} of the XRONOS package, adopting  the start time of the observation as reference time   and a resolution of the period search of $10^{-6}$ s. We explored  around a period  of 0.591116 s, estimated using eq.  (2) in \cite{Mazzola_19}, and subsequently, we fitted the peak of the corresponding  $\chi^2$ curve with a Gaussian function.  We assumed  that the centroid of the Gaussian function is the best estimate of the spin period. We found that the spin period is 0.591124781  s, the $\chi^2$ peak associated with the best period is 135 (see the left panel in Fig. \ref{fig:efsearch}), and the probability of obtaining a $\chi^2$  value greater than or equal to the $\chi^2$   peak by chance, having   15    degrees of freedom, is $2.2 \times 10^{-21}$   for a single trial. Considering the 1000 trials in our research, we expect almost $ 2.2 \times 10^{-18}$    periods with a $\chi^2$ value greater than or equal to $\chi^2$  peak. This implies a detection significance with $\sigma \simeq 8.7$.  
 
At the light of this result we searched for the spin  period   in the {\it NuSTAR} obsid. 30301009002 discussed by \cite{Anitra_21}. The FPMA light curve was rebinned at 0.01 s and we selected the events in the 5-40 keV energy range.  After applying the barycentric correction to the event arrival times  we corrected them   for the orbital motion adopting the eclipse arrival time of $T_{ecl} = 58234.1536(5)$  MJD/TDB \citep[see][]{Anitra_21}  and the orbital period at the time of the observation. Using  the ftool {\tt efsearch}, we explored around a period P$_{spin}$  of 0.591472 s  obtaining a  spin period of  0.59145871   s. The $\chi^2$ peak associated with the best period is 59 (see the right panel in Fig. \ref{fig:efsearch}), and the probability of obtaining a $\chi^2$  value greater than or equal to the $\chi^2$   peak by chance,    with 15 degree of freedom, is $3.7 \times 10^{-7}$  for a single trial. Considering the 1000 trials in our research, we expect almost $3.7 \times  10^{-4}$    periods with a $\chi^2$ value greater than or equal to the $\chi^2$  peak. This implies a detection significance with $\sigma \simeq 3.4$.   

Finally, we searched for periodicity in the NICER data without success. We searched  in the 2-10 keV  and   5-10 keV energy band because  the pulse amplitude increases  with increasing energy, it  is less than 0.5\% between 2 and 6 keV and  less than  1\%  between  6  and 10 keV \citep[see][]{jonker_01}. We believe that the low effective area of NICER above 5 keV and the short exposure time prevent us from finding a statistically significant signal.

We refined the  long-term spin period   derivative  using    the  spin period best-values,   which span 25 yrs,
shown in Tab. \ref{tab:pspin}. We fitted the P$_{spin}$ values with a linear function  (see the top panel of Fig. \ref{fig:pspin}) and associated the post-fit errors to the parameters. We found 
   a dependence of P$_{spin}$ on time given by:   
\begin{equation}
\label{eq:pspin}
\centering
        P_{\rm spin}(t)=0.592772(5) - 2.64(2) \times 10^{-12} (t-52500) \times 86400 {\; {\rm s}},  
\end{equation} 
where the time $t$ is given in MJD and the spin period derivative is $\Dot{P}_{spin} = -2.64(2) \times 10^{-12}$ s s$^{-1}$ ($\dot{\nu}=7.55(4) \times 10^{-12}$ Hz s$^{-1}$)  confirming that X1822-371 is spinning up.
 Propagating the errors on the ${P}_{spin}$  from eq.2 and using the eclipse arrival times from the two NuSTAR observations, we obtained the errors on the ${P}_{spin}$. We found that  ${P}_{spin}=0.59112(2)$ s and ${P}_{spin}=0.59144(2)$ s for the NuSTAR observations taken in 2022 and 2017, respectively.

   Fluctuations  are present in the data  (see bottom panel in Fig. \ref{fig:pspin}) as already discussed by \cite{Bak17} although the decreasing trend is evident.   We show the pulse profiles  in Fig. \ref{fig:profile}. By fitting with a sinusoidal function we found that the pulse amplitude is 1.8\% for both the observations. Finally, we note that  the pulse profiles 
are not perfectly sinusoidal and their shape is similar to that shown by  \cite{jonker_01} for  the energy band 9.4-22.7 keV.  

\subsection{The  NICER  spectrum}

\begin{table*}
    \centering
   \scriptsize
        \caption{Adopted models to fit the  NICER  spectrum}
    \begin{tabular}{clcccccc}
      \toprule
        Model & components & $\chi^2$(d.o.f.) & $\Delta \chi^2$$^a$  & significance$^b$  & $E$$^c$   & $\sigma$$^d$  & Strength$^e$\\
              &      &  &  &  & (keV) & (keV) &$(\times 10^{-2})$\\  
        \midrule
    A     & \texttt{TBabs*(Comptb+gauss+gauss)} & 453(144)&--&--&--&--&--\\
     B    & \texttt{TBabs*(bbodyrad+Comptb+gauss+gauss)} & 258(142)& 172.9 & $12$ &$0.727\pm0.015$ &$0.11\pm0.03$ & $7^{+3}_{-2} $\\
     C    & \texttt{TBfeo*(bbodyrad+Comptb+gauss+gauss)} & 113(140)& 27.7 &3.1 &$0.73^{+0.03}_{-0.05}$ &$0.10\pm0.03$ & $6^{+6}_{-3} $ \\
     D    & \texttt{TBabs*edge*edge*(bbodyrad+Comptb+gauss+gauss)} & 99(140) & 15.1 & 0.1 &$0.650^{+0.118}_{-0.012}$& $0.05^{+0.07}_{-0.05}$ &$3.3^{+1.5}_{-2.4}$ \\
     E    & \texttt{TBpcf*TBabs*(ga$_{OK}$+ga$_{Ne9}$+Comptb+gauss+gauss)}$^f$  & 116(138)&-- &--&--&--& -- \\
     F    & \texttt{TBpcf*TBabs*(ga$_{OK}$+ga$_{Ne9}$+Comptb+gauss+gauss)}$^g$  &  136(140)& 48.2 &5.3&$0.74 \pm 0.03$& $0.08\pm0.04$& $2.4^{+1.0 }_{-0.7}$\\
     G & \texttt{gabs*TBabs*(bbodyrad+Comptb+gauss+gauss)}& 
        85(139) &--&-- \\
      \bottomrule 
          \end{tabular}
              \begin{tablenotes}
\item[]      $^a$Obtained adding to the model a \texttt{gabs} component at 0.7 keV. 
           $^b$Statistical significance for the addition of the \texttt{gabs} component at 0.7 keV. 
     $^{c,d,e}$ Best-fit values of the energy, width and strength of the \texttt{gabs component}, errors are at 90\% confidence level.  
        $^f$The widths of the neutral O and \ion{Ne}{ix} emission lines  were left free to vary.
        $^g$The widths of the neutral O and \ion{Ne}{ix} emission lines were kept fixed at 5 eV.
    \end{tablenotes}
   \label{tab:stat_nicer}
\end{table*}

\begin{figure}[]
   \centering
    \includegraphics[scale=.65]{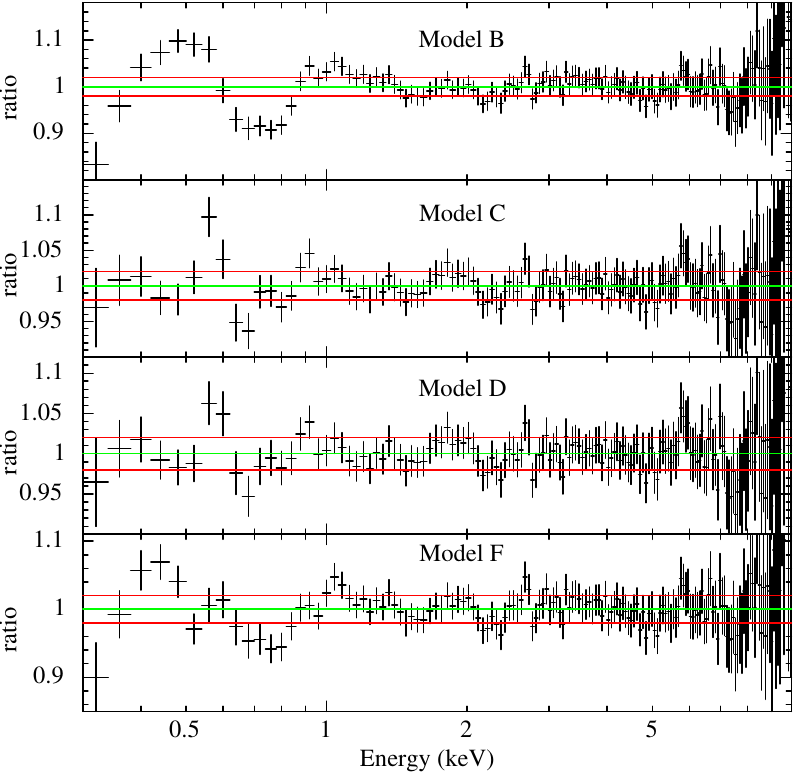}
    \caption{0.3-10 keV ratios for  Models  B, C, D and F. The systematic error is that suggested by the  NICER  calibration team. The red horizontal lines indicate a deviation of the model from the data of 2\%.}
   \label{fig:residualsfromBtoI}
\end{figure}
\begin{figure}[]
   \centering
    \includegraphics[scale=.65]{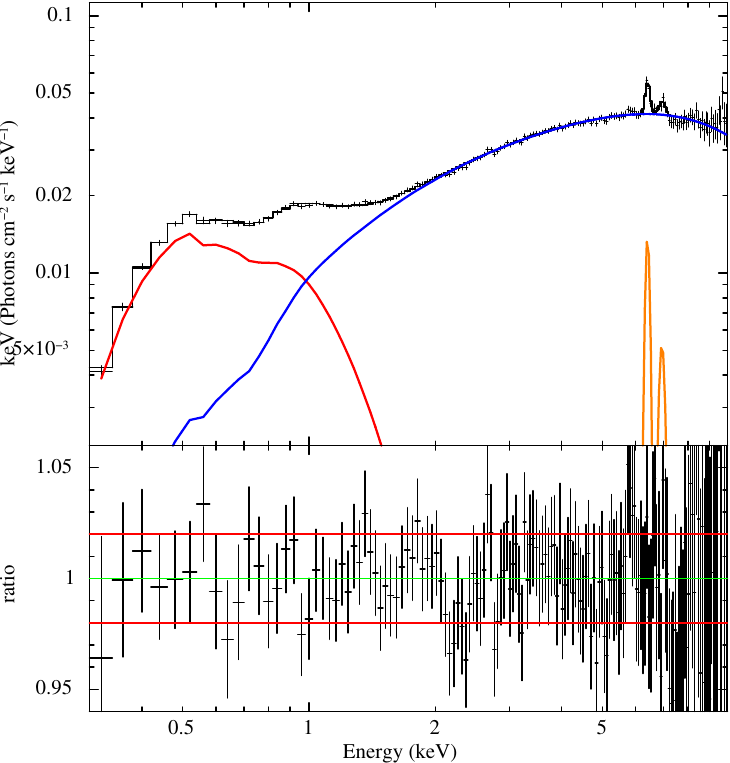}
    \caption{0.3-10 keV unfolded spectrum using Models G (top panel); the blackbody, the Comptonized component and the emission lines at 6.4 keV and 6.96 keV are shown with the red, blue and orange colors, respectively.  The corresponding data-to-model ratio is in  bottom panel.}
   \label{fig:modelGsyscal}
\end{figure} 
Initially, we fitted only the NICER spectrum in the 0.3-10 keV energy band with XSPEC  v12.13.0c. We adopted the {\tt TBabs} component  accounting for  the ISM abundances   \citep{Wilms00}  and set  the photoelectric cross section reported  by  \cite{Verner_96}. We adopted a model  composed of  a Comptonized component  \citep[{\tt Comptb} in XSPEC,][]{farinelli_08} and   two Gaussian emission lines at 6.4 keV and 6.96 keV.
We kept fixed the parameter $logA$ at the value of 8 in the  {\tt Comptb} component. The 
parameter $logA$ is called the illumination factor, where $1/(1 + A) $ is the fraction of the seed-photon radiation directly seen by the observer, whereas $A/(1 + A) $ is the fraction upscattered by the Compton cloud. By fixing  $logA=8$ we are assuming that all the seed-photons are upscattered.  Hence, the adopted model is {\tt Model\;A: TBabs*(Comptb+gauss+gauss)};
this simple model does not give a good fit of the spectrum;
we found a $\chi^2$(d.o.f.) value of 453(144). To improve the fit  we added a blackbody component ({\tt bbodyrad} in XSPEC) to the model ({\tt Model\;B: TBabs*(bbodyrad+Comptb+gauss+gauss)}) finding a $\chi^2$(d.o.f.) value of 258(142). The fit is still statistically unsatisfactory,  since large residuals are present below 1 keV, where the data-to-model ratio is larger than 5\% (see top panel in \autoref{fig:residualsfromBtoI}).  

A possible explanation for those residuals below 1 keV is that there might be abundances of oxygen and/or iron deviating from solar values. Therefore, we replaced the photoelectric absorption model \texttt{Tbabs} with the \texttt{Tbfeo} model, allowing the oxygen and iron abundances to vary\footnote{\url{https://heasarc.gsfc.nasa.gov/docs/nicer/data_analysis/workshops/NICER-CalStatus-Markwardt-2021.pdf}}. We fitted the  spectrum adopting   {\tt Model\;C: TBfeo*(bbodyrad+Comptb+gauss+gauss)}. We obtained 
 a $\chi^2$(d.o.f.) value of 113(140), the abundances of oxygen and iron were $3.6 \pm 0.4$  and $> 10$ times the solar abundances, respectively.
 Regardless of the plausibility of the abundance values associated with oxygen and iron, we still notice that the ratio exhibits residuals larger than 5\% around 0.7 keV (see the second panel from the top in \autoref{fig:residualsfromBtoI}).

We  replaced the \texttt{Tbfeo}   component in the model with \texttt{Tbabs}  and added two absorption edges (\texttt{edge} in XSPEC) with threshold energies fixed at 0.56 keV and 0.71 keV. 
The two absorption edges are associated with the neutral K-shell O and neutral L-shell Fe.  This choice implies that the neutral iron abundance estimated by the \texttt{edge} at 0.71 keV  is different from that estimated by the neutral iron K-shell edge at 7.1 keV  with \texttt{Tbabs} component, making the model non-self-consistent.
The sole purpose of this model is to address the residuals below 1 keV. We fitted the spectrum with {\tt Model\;D: TBabs*edge*edge(bbodyrad+Comptb+gauss+gauss)}.
We found a $\chi^2$(d.o.f.) value of 99(140).
The ratio corresponding to this model is shown in the third panel from the top  in \autoref{fig:residualsfromBtoI}. We observe that close to 0.7 keV the model deviates from the data by 5\%. Since the introduced systematic error is less than 2\%, we conclude that this model, besides being non-self-consistent, fails in its purpose of accurately modeling the spectrum below 1 keV.

To take into account   a potential misrepresentation of the Solar Wind Charge Exchange (SWCX) background during the observation 
we fitted the spectrum adopting  {\tt Model\;E: TBpcf*TBabs*(ga$_{OK}$+ga$_{Ne9}$+Comptb+gauss+gauss)}, 
where the components \texttt{ ga$_{OK}$} and
\texttt{ ga$_{Ne9}$} are two Gaussian emission lines with energy fixed at 0.53 keV and 0.92 keV and associated with neutral oxygen and \ion{Ne}{ix}, respectively. 
 Initially,  we  let the widths of the two lines free to vary   and
obtained a $\chi^2$(d.o.f.) value of 116(138). However,  the best-fit values of the widths of the K-shell O and the \ion{Ne}{ix} lines  are $60 \pm 15$ eV and $280^{+100}_{-50}$ eV; the broadening of the two  lines suggests that the Gaussian components try to fit the  continuum emission.   In fact,  SWCX lines are typically modeled as delta functions \cite[see e.g.,][]{Fujimoto07}, so this model must be rejected. Then, we fitted the spectrum keeping fixed  the widths of the \texttt{ ga$_{OK}$} and
\texttt{ ga$_{Ne9}$}  components  at 5 eV (hereafter {\tt Model\;F}). We found a $\chi^2$(d.o.f.) value of 136(140) but large residuals persist below 1 keV as shown in the bottom panel of \autoref{fig:residualsfromBtoI}.

Moreover, we investigated  the possibility that the residuals below 1 keV are due to a poor modeling of the \texttt{SCORPEON} background spectrum. Therefore, we extracted the source spectrum and the background model (using the \texttt{nicerl3-spect} tool and setting \texttt{bkgformat=script}) to simultaneously fit the source spectral model and the \texttt{SCORPEON}  background model\footnote{\url{https://heasarc.gsfc.nasa.gov/docs/nicer/analysis_threads/scorpeon-xspec/}}. 
To fit the source spectrum and background together we adopted the pgstat statistic.
However, the fit in the  0.3-15 keV energy range leaves the broad residuals below 1 keV unchanged with a ratio larger than 4\%, suggesting that the residuals observed at 0.7 keV are not related to a mismodeling of the background. 

Before proceeding with the investigation of the X 1822-371 spectrum below 1 keV, we note that the models from C to F (summarized in \autoref{tab:stat_nicer}) show a reduced $\chi^2$  less than 1 but the corresponding data-to-model ratios exceed 5\% below 1 keV (see \autoref{fig:residualsfromBtoI}). The reduced $\chi^2$  values should not lead to the conclusion that we   found a good fit because they are influenced by the systematic error we added to the spectrum. On the other hand, ratios exceeding 2\% below 1 keV indicate the need to add an additional component to obtain a good fit of    the spectrum below 1 keV\footnote{\url{https://heasarc.gsfc.nasa.gov/docs/nicer/analysis_threads/plot-ratio/}}.
 
Analyzing a \textit{XMM-Newton} observation of X 1822-371, \cite{iaria_15} suggested the presence of a cyclotron line at 0.7 keV, and  we decided to investigate  this scenario. We added to  \texttt{Model B} a   {\tt gabs} component 
at 0.7 keV to take into account the possible presence of a cyclotron line. (hereafter {\tt Model\;G: gabs*TBabs*(bbodyrad+Comptb+gauss+gauss)}).
We obtained 
a $\chi^2$(d.o.f.) value of 85(139) 
with a $\Delta \chi^2$ of 172.9 with respect to {\tt Model\;B} and the residuals below 1 keV disappear. We show the unfolded model and the corresponding ratio in \autoref{fig:modelGsyscal},   the best fit values of the parameters associated with this model are shown  in the third column of \autoref{Tab:model_G}.

\begin{figure}[]
   \centering
    \includegraphics[scale=.65]{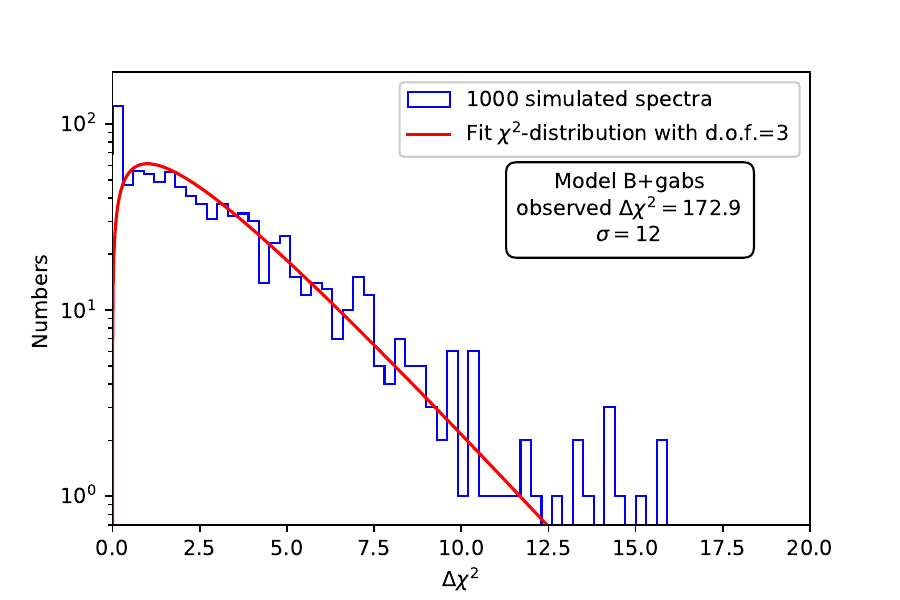}
    \caption{Monte-Carlo simulations results for testing the \texttt{gabs} significance for  \texttt{Model B}.   The blue histogram shows $\Delta \chi^2$ values obtained in the simulations, and the red curve is a $ \chi^2$  distribution with 3 d.o.f..  The  $\Delta \chi^2=172.9$   attained in actual data   is significantly higher   than values attained in simulations (by $\sim 12 \sigma$).}
   \label{fig:histo}
\end{figure}

To verify whether the {\tt gabs}   component at 0.7 keV is statistically significant we used Monte-Carlo simulations as shown by \cite{Bharelao_15}. We adopted {\tt Model\;B} as our null hypothesis and simulated 1000 fake spectra. Then we fitted each fake spectrum with {\tt Model\;B} and {\tt Model\;B} plus 
{\tt gabs}   (i.e. {\tt Model\;G}) annotating the $\Delta \chi^2$ value obtained for the addition of   {\tt gabs}. 
Since the cyclotron line adds three free parameters, we expect that the histogram of $\Delta \chi^2$  values  follows a $\chi^2$ distribution with three degrees of freedom. We show the results in \autoref{fig:histo}.
The highest $\Delta \chi^2$    obtained in our simulation is 15.8, significantly lower than $\Delta \chi^2=172.9$ obtained in real data. From our simulation the probability to observe a $\Delta \chi^2$ of 172.9 by chance  is $7.8 \times 10^{-35}$ that corresponds to a significance of $\sim 12\, \sigma$.  

We repeated the same procedure for Model C, D    and F.  We show the 
observed $\Delta \chi^2$ due to the addition of the   \texttt{gabs} component and its statistical significance  in the fourth and fifth column of \autoref{tab:stat_nicer}.
The addition of the \texttt{gabs} component to the tested models is always larger than 3 sigmas except for \texttt{Model D} for which we find the significance is only 0.1 $\sigma$.
So, except for \texttt{Model D},
the addition of a \texttt{gabs} component significantly improves the quality of the fit.

\begin{table}
\scriptsize  
    \centering
            \caption{Best-fit values of the parameters associated with \texttt{Model G}}
    \begin{tabular}{llcc}
    \toprule
          Component& Parameter & \multicolumn{2}{c}{\texttt{Model G}}\\
         &  & \texttt{syst=CALDB} & \texttt{syst=1\%}\\
         \midrule
      \texttt{Tbabs}   & N$_H$ ($\times 10^{20}$) & $6.7\pm1.0$& $6.3^{+1.1}_{-0.9}$\\ 
      \texttt{gabs}   & E (keV) & $0.727\pm0.015$&$0.728\pm0.013$ \\
         & $\sigma$ (keV) & $0.11\pm0.03$ & $0.11\pm0.02$\\
         & Strength  ($\times 10^{-2}$)& $7^{+3}_{-2}$ &$6.3^{+2.0}_{-1.3}$   \\ 
    \texttt{bbodyrad}     & kT (keV) &  $0.179\pm0.007$  &$0.181\pm0.007$\\
         & N$_{bb}$  & $2600^{+700}_{-500}$ 
         &$2400^{+600}_{-400}$ \\ 
   \texttt{Comptb}      & KT$_s$ (keV)& 
   $0.71\pm0.09$& $0.72\pm0.08$\\
         & $\alpha$ & $0.28\pm0.04$& $0.29\pm0.03$\\
         & KT$_e$ (keV) & $2.8\pm0.3$ &$2.8\pm0.2$\\
         & $\log(A)$ & 8 (fixed)& 8 (fixed) \\
         &  N$_{comptb}$ ($\times 10^{-2}$) & $3.4\pm 0.5$ & $3.4\pm 0.4$\\ 
  \texttt{gauss}       & E (keV) & $6.41 \pm0.03$& $6.41 \pm0.02$\\
         & $\sigma$ (eV) & $80\pm40$ & $80\pm40$\\
         & I ($\times 10^{-4}$) & $4.5\pm1.1$ & $4.5\pm1.0$ \\ 
 \texttt{gauss}         & E (keV) & $6.93^{+0.08}_{-0.17}$ &$6.91^{+0.09}_{-0.16}$\\
         & $\sigma$ (eV) & $<250$&
         $140^{+130}_{-90}$\\
         & I ($\times 10^{-4}$)& $2.3^{+1.3}_{-1.2}$&
         $2.4^{+1.3}_{-1.0}$ \\ 
         & $\chi^2$(dof) & 85(139) & 117(138)\\
         \bottomrule
    \end{tabular}
        \begin{tablenotes}
\item[] Best-fit values adopting the systematic error suggested by the NICER   calibration team (third column) and  a systematic error of 1\% (fourth column). The  errors are at 90\% c.l. 
        \end{tablenotes}
    \label{Tab:model_G}
\end{table}

\subsection{Combined fit of the  NICER  spectra}
    \begin{figure*}[]
   \centering
    \includegraphics[scale=.6]{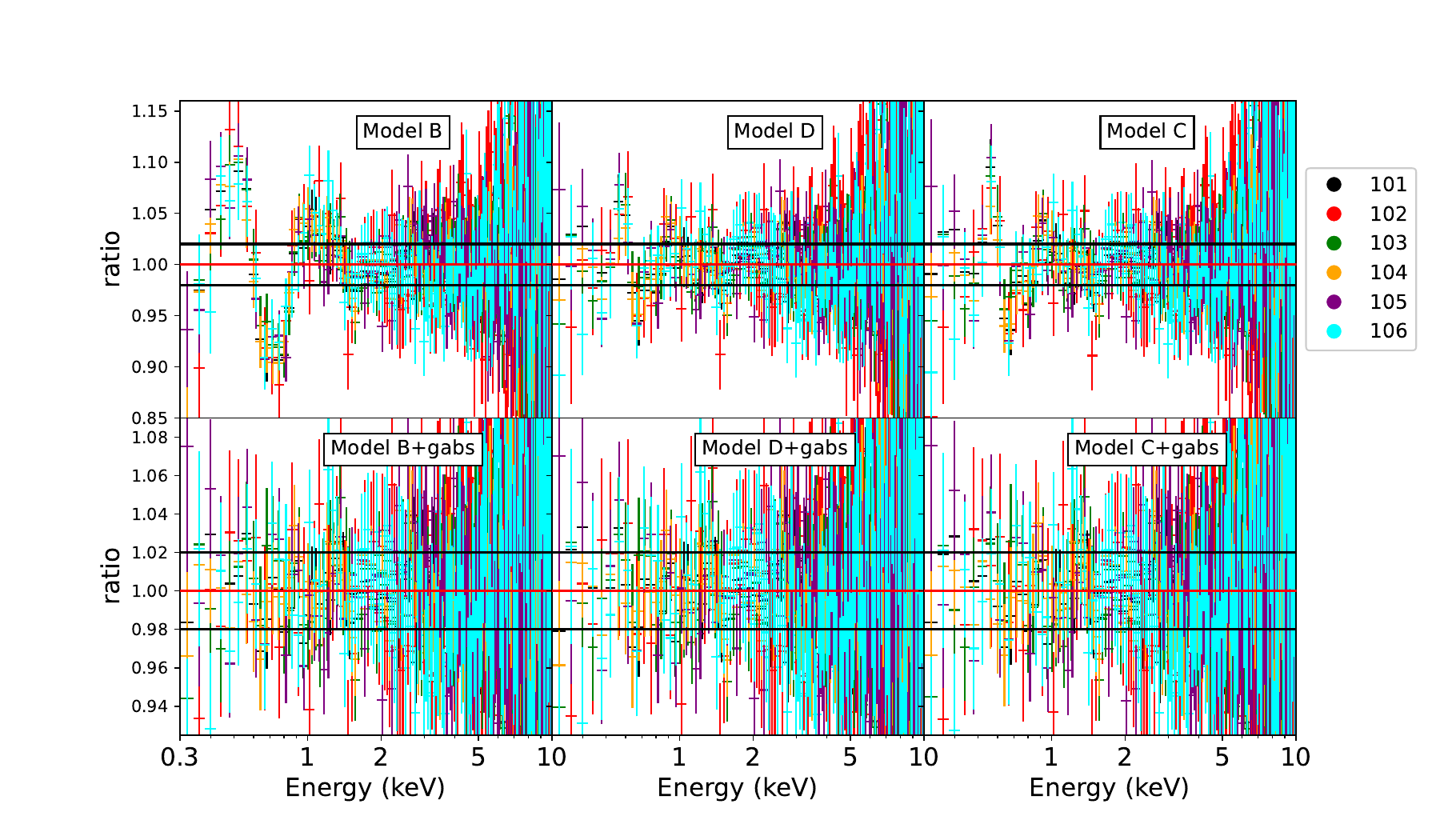}
    \caption{Data-to-model ratio of the six  NICER  (0.3-10 keV) spectra with respect to the different adopted models, with or 
without the gabs component. The 
adopted colors for each spectrum are shown in the legend in which we indicate the last three digits of the corresponding observation ObsId.  In the top panels we show the ratio corresponding to \texttt{Model B},  \texttt{Model C}, and  \texttt{Model D}, respectively. In the bottom panels we show the ratio of the same models including a \texttt{gabs} component at 0.7 keV.   The black horizontal  lines indicate a deviation of the model from the data corresponding to 2\%.  }
   \label{fig:nicer_all_res}
\end{figure*} 
 \begin{figure*}[]
   \centering
    \includegraphics[scale=.39]{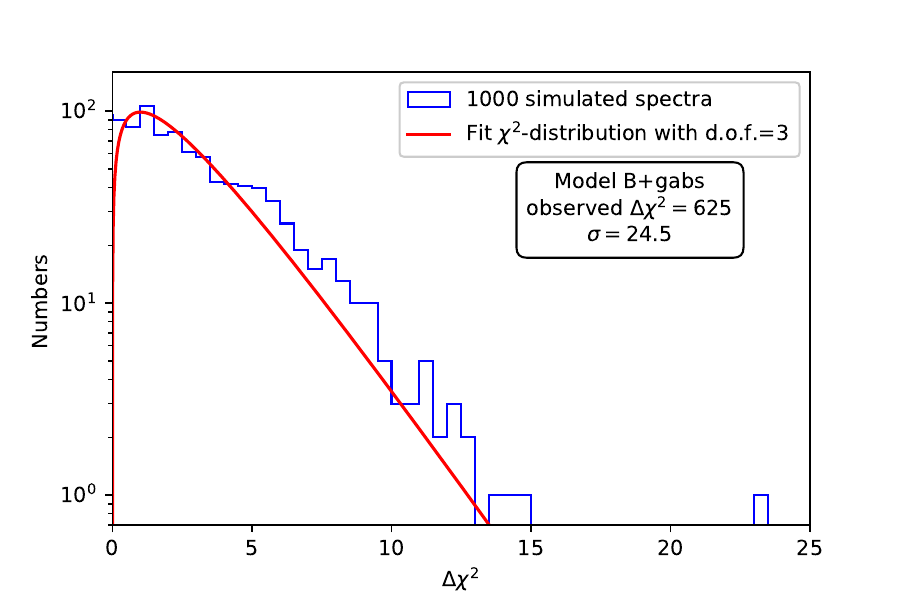}
    \includegraphics[scale=.39]{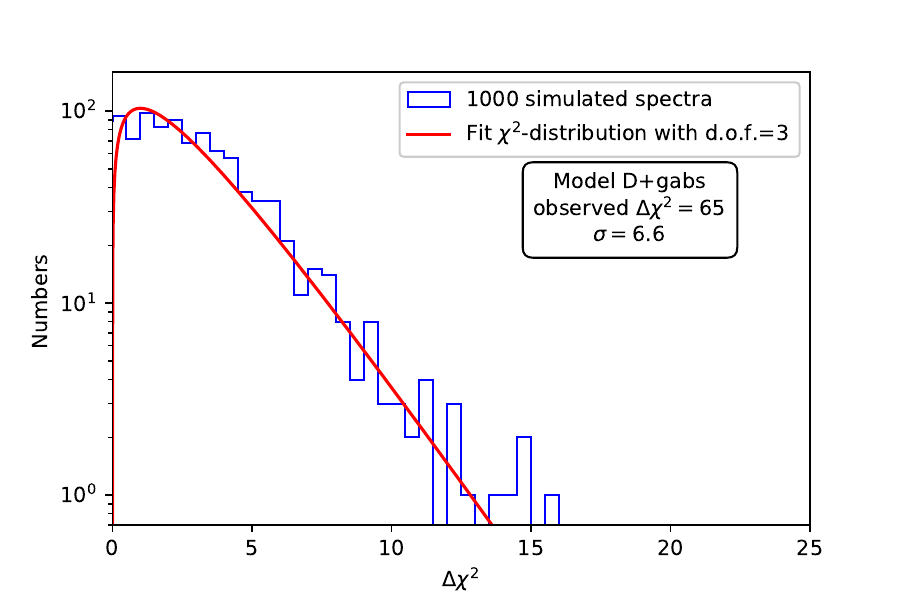}
    \includegraphics[scale=.39]{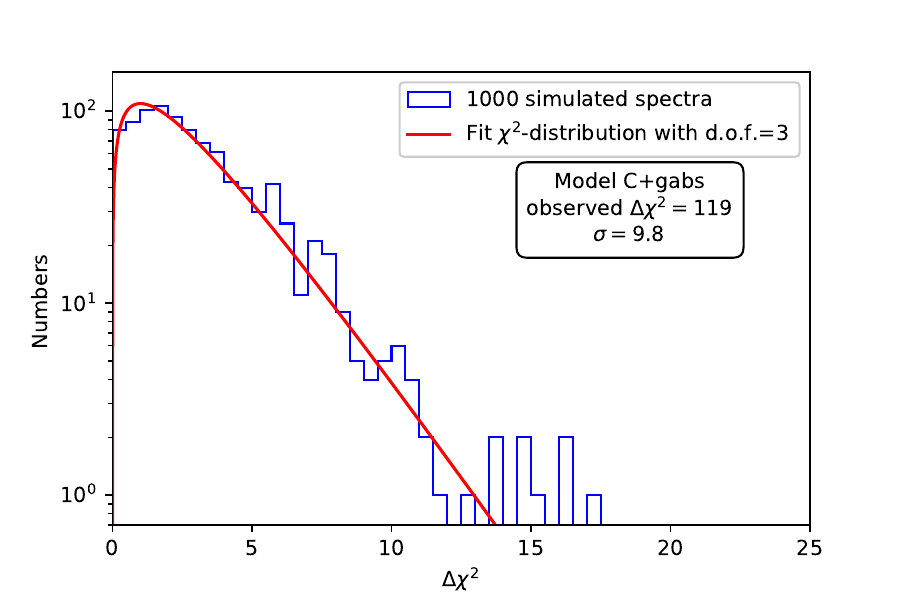}
    \caption{Monte-Carlo simulation results to assess the statistical significance of adding the gabs component into  \texttt{Model B}, \texttt{Model C}, and  \texttt{Model D}.    The addition of the \texttt{gabs} component is significant at  24.5 $\sigma$,  6.6 $\sigma$, and 9.8 $\sigma$ in \texttt{Model B}, \texttt{Model D}, and  \texttt{Model C}, respectively.}
   \label{fig:histo_ni_all}
\end{figure*} 

To further investigate the possible presence of   overabundances  of neutral oxygen and neutral iron with respect to the solar-abundances values, that is   \texttt{Model C} and \texttt{Model D}, we combined the aforementioned  NICER   spectrum with five additional  NICER  spectra obtained from five different pointings of the source.

The five  NICER  observations (obsid. 
from 5202780102 to 5202780106)
have an exposure time of 1.4 ks, 4.3 ks, 5.9 ks, 2.4 ks and 1.7 ks, respectively.  
The observations were conducted from June 23 to June 26, 2022.
During the observation the Focal Plane Module (FPM) 63   was noisy, and during the screening of the data the tool nicerl2 rejected all the events collected by FPM 63.  Moreover, in order to produce  the spectrum we removed FPM 14 and 34 using  {\tt nicerl3-spect} tool because of increased detector noise. 

The source spectrum and the {\tt scorpeon} background file  were  extracted using the tool  {\tt nicerl3-spect} and setting the options {\tt bkgformat=file}.  We added the systematic error suggested by the  NICER  calibration team   to the spectrum  and grouped the data  using the ftool {\tt ftgrouppha} by applying an optimal rebinning {\tt grouptype=optmin} to have at least 25 counts per energy bin  \citep[{\tt groupscale=25},][]{Kaastra_16}.   

To fit the six spectra  we included to the analyzed models a constant component (\texttt{const}) to account for flux variations of the source in different observations. Additionally, we left the blackbody normalization and the temperature of the seed photons (kT$_s$) free to vary for all six spectra. Finally, we added an edge  at 2.1 keV to address the calibration feature at 2  keV  present in the NICER spectra (see \autoref{fig:modelGsyscal}). 

Initially, we fitted the six spectra using Model B.  We found 
a $\chi^2$(d.o.f.) of 1219(844). Below 1 keV, the model deviates from the data by 10\%, as shown in the top-left panel of  
\autoref{fig:nicer_all_res}. By adding 
a \texttt{gabs} component at 0.7 keV we obtained a $\chi^2$(d.o.f.) of 594(841) and 
a $\Delta \chi^2$ of 625. We show the ratio corresponding to this model in the bottom-left panel of \autoref{fig:nicer_all_res}, the residuals below 1 keV disappear. 

To verify whether the {\tt gabs}   component at 0.7 keV is statistically significant we used Monte-Carlo simulations. We adopted {\tt Model\;B} as our null hypothesis and simulated 1000 fake spectra. We show our simulations is the left panel of \autoref{fig:histo_ni_all}, the largest value of $\Delta \chi^2$ obtained from the simulations is 23.2 that is  significantly smaller than $\Delta \chi^2=625$ obtained in real data.
The probability to observe a $\Delta \chi^2$ of 625 by chance is $1.6\times 10^{-132}$ corresponding to 24.5 $\sigma$. The best-fit parameters of the  {\tt gabs}   component are $E=0.726\pm0.008$ keV, $\sigma=0.106\pm0.012$ keV and Strength$=0.064^{+0.011}_{-0.009}$. We conclude that the addition of the {\tt gabs}   component  to \texttt{Model B} is highly significant. 

By fitting the data using \texttt{Model D}
 We found 
a $\chi^2$(d.o.f.) of 657(842). 
The depths of the 
edges at 0.56 keV and 0.71 keV are $0.31\pm 0.03$ and $0.20 \pm 0.02$, respectively. Below 1 keV, the model deviates from the data by at least  5\%, as shown in the top-center panel of  
\autoref{fig:nicer_all_res}. Adding 
a \texttt{gabs} component at 0.7 keV we obtained a $\chi^2$(d.o.f.) of 592(839) and 
a $\Delta \chi^2$ of 65. We show the ratio corresponding to this model in the bottom-center panel of \autoref{fig:nicer_all_res}, the residuals below 1 keV disappear.  
The depth of the 
edge at 0.56 keV and 0.71 keV are $<0.17$ and $<0.12$, respectively. The best-fit parameters of the  {\tt gabs}   component are $E=0.74^{+0.02}_{-0.03}$ keV, $\sigma=0.09\pm0.02$ keV and Strength$=0.04\pm0.02$.

Similarly, we assessed the statistical significance of incorporating the  {\tt gabs}   component in this instance.
 We adopted {\tt Model\;D} as our null hypothesis and simulated 1000 fake spectra. We show our simulations in the center panel of \autoref{fig:histo_ni_all}, the largest value of $\Delta \chi^2$ obtained from the simulations is 15.7 that is smaller than $\Delta \chi^2=65$ obtained in real data.
The probability to observe a $\Delta \chi^2$ of 65 by chance is $2.1\times 10^{-11}$ corresponding to 6.6 $\sigma$.   We conclude that the addition of the {\tt gabs}   component  to \texttt{Model D} is highly significant.

Finally, we fitted the six  NICER  spectra adopting  {\tt Model\;C}, where we adopted the \texttt{Tbfeo} component leaving free to vary the O and Fe abundance. We found a $\chi^2$(d.o.f.) of 713(842); the abundances of O and Fe are  $3.4 \pm0.2$ and $11.5\pm1.3$ times the corresponding solar-abundance values.  
However, the ratio  in the top-right panel of \autoref{fig:nicer_all_res} shows that the model deviates from data by at least 5\% below 1 keV.
By adding to the model a \texttt{gabs} component at 0.7 keV we obtained a $\chi^2$(d.o.f.) of 594(839) and 
a $\Delta \chi^2$ of 119. We show the ratio corresponding to this model in the bottom-right panel of \autoref{fig:nicer_all_res}, the residuals below 1 keV disappear. The abundances of O and Fe are  $0.8 \pm0.5$ and $0.3^{+3.3}_{-0.3}$  times the corresponding solar-abundance values.  
The best-fit parameters of the  {\tt gabs}   component are $E=0.726^{+0.008}_{-0.022}$ keV, $\sigma=0.109^{+0.021}_{-0.014}$ keV and Strength$=0.07\pm0.03$.

Likewise, we evaluated the statistical significance associated with the inclusion of the {\tt gabs}  component in this   case.
 We adopted {\tt Model\;C} as our null hypothesis and simulated 1000 fake spectra. We show our simulations is the right panel of \autoref{fig:histo_ni_all}, the largest value of $\Delta \chi^2$ obtained from the simulations is 17.2 that is smaller than $\Delta \chi^2=119$ obtained in real data.
The probability to observe a $\Delta \chi^2$ of 119 by chance is $5.7\times 10^{-23}$ corresponding to 9.8 $\sigma$.   We conclude that the addition of the {\tt gabs}   component  to \texttt{Model C}, is highly significant.
So, the inclusion of the {\tt gabs}   component is always necessary in models \texttt{B}, \texttt{C}, and \texttt{D}, as the addition of this component exhibits a significance exceeding 5 sigma for all the three examined models.

\subsection{Combined fit of   the  NICER  spectrum with two \textit{XMM/EPIC-pn} spectra}
\begin{figure*}[!htbp]
   \centering
    \includegraphics[scale=.58]{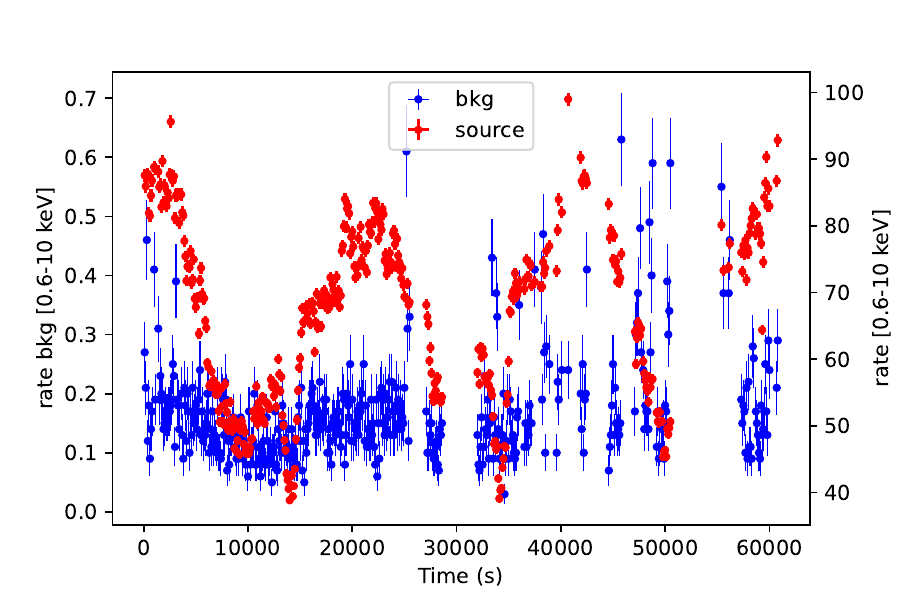}
    \includegraphics[scale=.58]{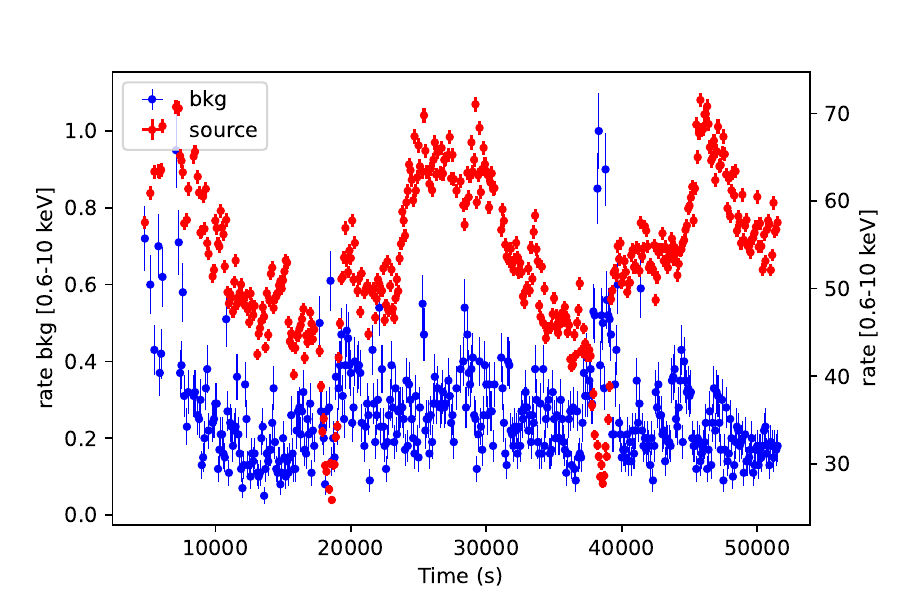}    
    \caption{ 0.6-10  keV EPIC-pn light curve (source in red, background in blue)  corresponding to the obsid. 0784820101 (left panel) and obsid. 0111230101 (right panel). Bin time 100 s.}
   \label{fig:lightcurve_078482010}
\end{figure*}
\begin{figure}[h]
   \centering
    \includegraphics[scale=.60]{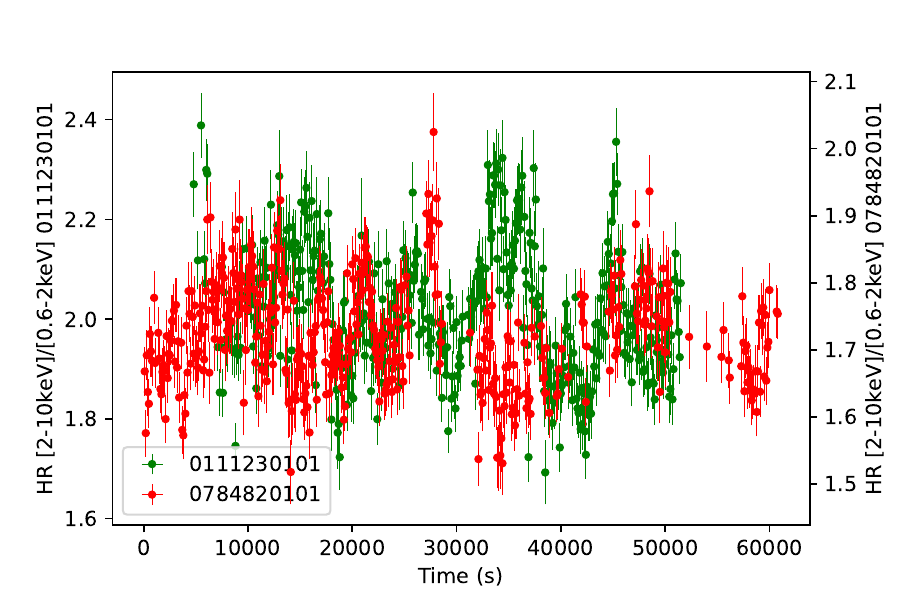}
    \caption{ EPIC-pn hardness ratios of the XMM observation made in 2001 (green color) and in 2017 (red color). The bin time is 100 s.}
   \label{fig:hr}
\end{figure}

To further enhance the robustness of our detection  of the cyclotron line at 0.7 keV and to better investigate whether there are two edges associated with the K-shell of neutral oxygen and the L-shell of neutral iron  below 1 keV, we reanalyzed two \textit{XMM/EPIC-pn}  spectra  combining them with the    NICER  spectrum.

The obsid. 0784820101 was already analyzed by \cite{Anitra_21}.  The authors analyzed the EPIC-pn spectrum in the 2-10 keV energy band suggesting calibration issues in the spectrum associated with the presence of a silicon edge at 1.8 keV. 
The source   was observed by the XMM-{\it Newton}  observatory on 2017 March 3  for a duration of 69 ks. During the observation the EPIC-pn camera was operating in Timing Mode. We reduced the \textit{XMM-Newton} data using the Science Analysis Software (SAS) v21.0.0. 
Initially, we extracted the EPIC-pn events between 10 and 12 keV to verify the presence of particle flaring background\footnote{\url{https://www.cosmos.esa.int/web/xmm-newton/sas-thread-epic-filterbackground}}. The light curve shows a strong flaring activity after the first 30 ks from the beginning of the observation. To mitigate the effects of particle flaring, we selected the time intervals with a rate lower than 1.5 c/s   in the energy range 10-12 keV. 
    \begin{figure*}[]
   \centering
    \includegraphics[scale=.6]{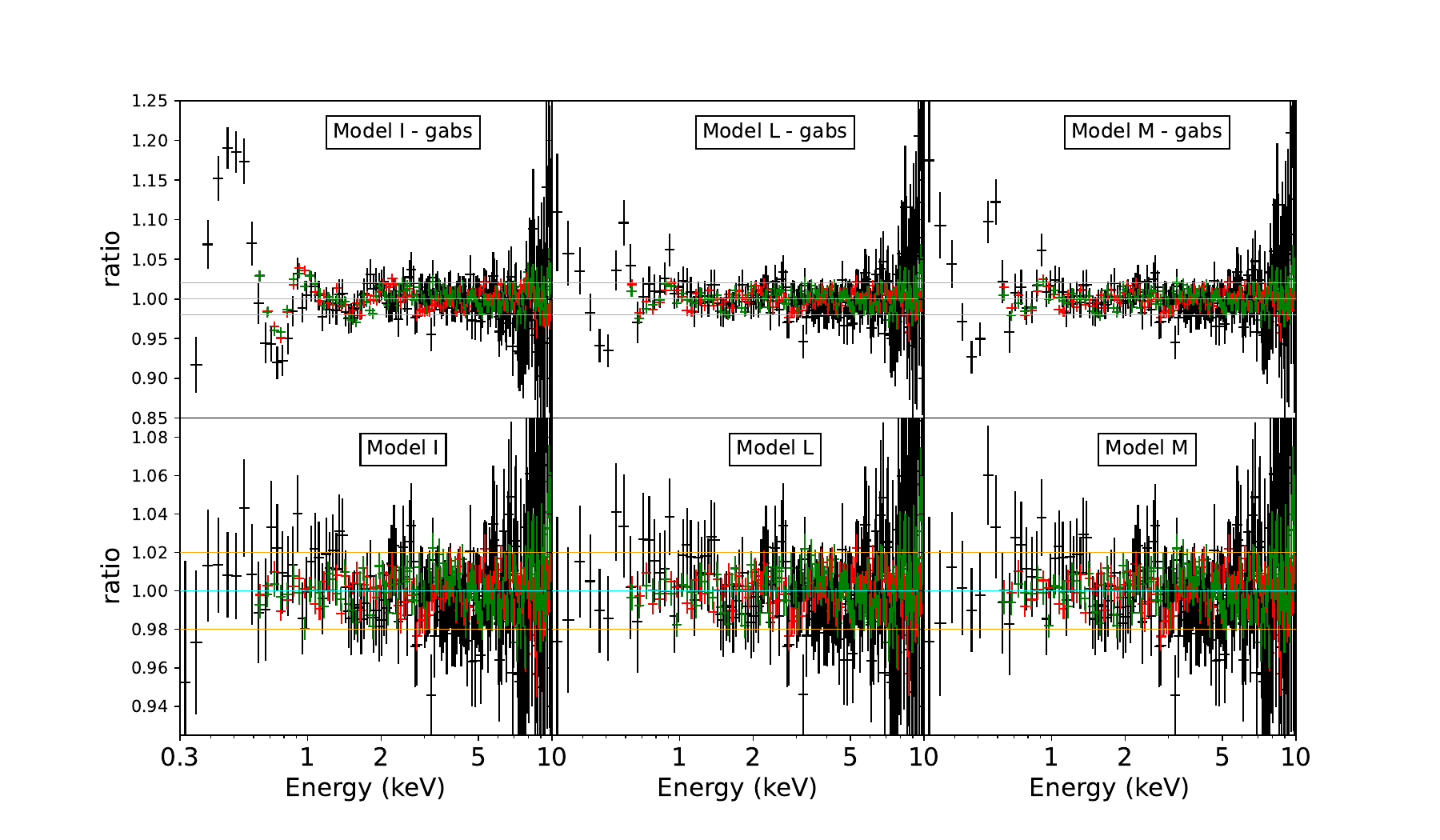}
    \caption{Ratios combining the  0.3-10 keV  NICER   (black),  PN1 (red) and 
    PN2 (green) spectrum.  In the top panels the residuals corresponding to \texttt{Model I},  \texttt{Model L}, and  \texttt{Model M}   removing the \texttt{gabs} component. In the bottom panels the residuals associated with the best-fit model shown in  \autoref{tab:compare_spectrum}. The orange horizontal  lines indicate a deviation of the model from the data larger than 2\%.  }
   \label{fig:nicer_epn07}
\end{figure*} 
 \begin{figure*}[]
   \centering
    \includegraphics[scale=.39]{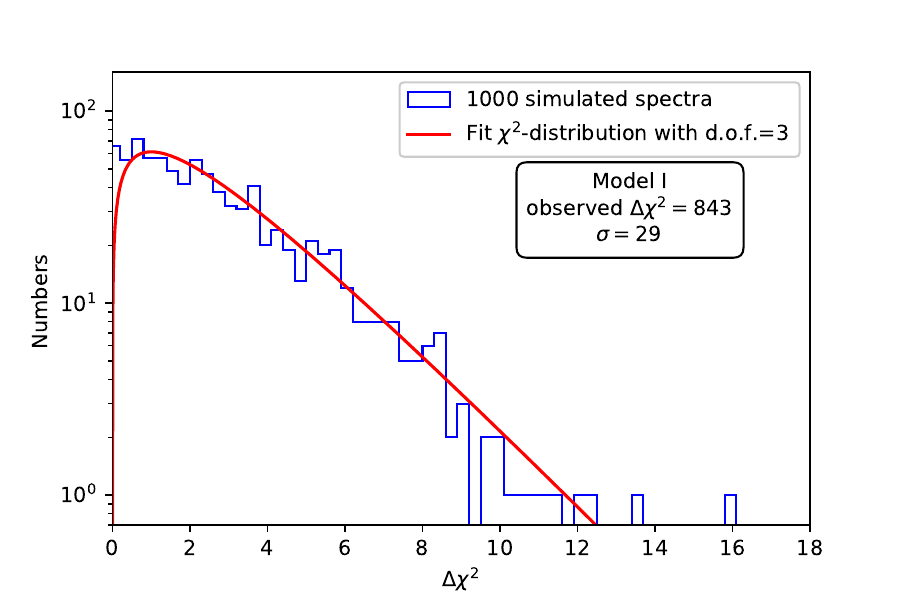}
    \includegraphics[scale=.39]{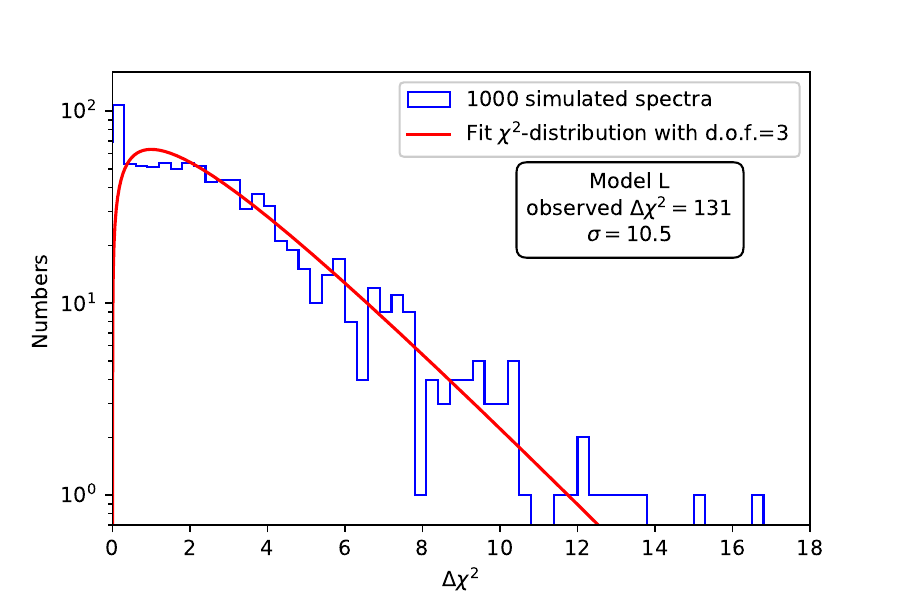}
    \includegraphics[scale=.39]{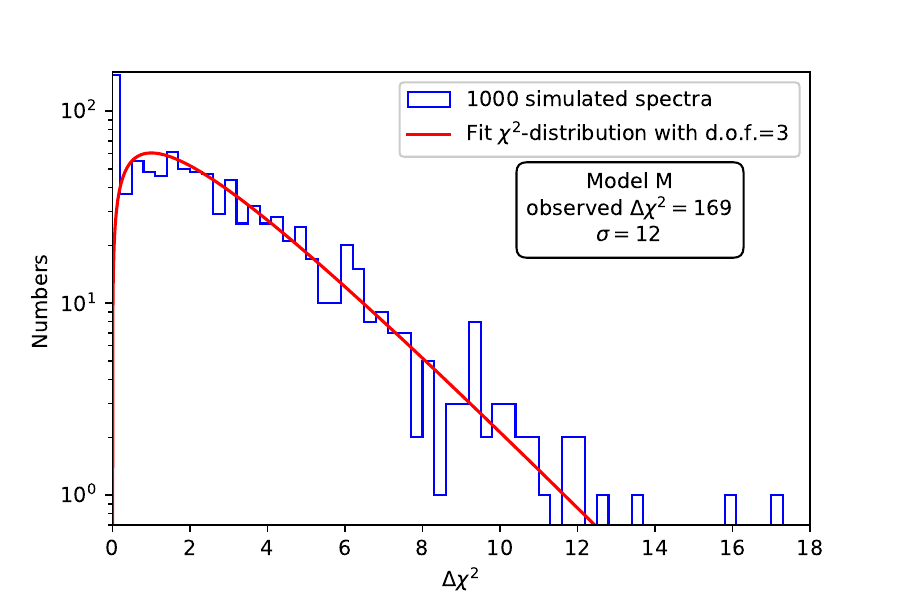}
    \caption{Monte-Carlo simulation results to assess the statistical significance of adding the gabs component into  \texttt{Model I}, \texttt{Model L}, and  \texttt{Model M}.     The addition of the \texttt{gabs} component is significant at  29 $\sigma$,  10.5 $\sigma$, and 12 $\sigma$ in \texttt{Model I}, \texttt{Model L}, and  \texttt{Model M}, respectively.}
   \label{fig:histo_ni_pn1_pn2}
\end{figure*}

We show the source and background EPIC-pn light curve in the 0.6-10 keV energy range cleaned from the particle flaring background in   \autoref{fig:lightcurve_078482010} (left panel). The source light curve clearly shows the orbital modulation and ranges between 40  c s$^{-1}$ and 90 c s$^{-1}$, two partial eclipses are present at 14000 s and 34000 s from the beginning of the observation.

The EPIC-pn spectrum and background were extracted selecting  RAWX between 26 and 47 and RAWX between 3 and 5, respectively. The EPIC-pn spectrum has an exposure time of 41 ks, it was   grouped  using the ftool {\tt ftgrouppha} by applying an optimal rebinning 
{\tt grouptype=optmin} to have at least 25 counts per energy bin  \citep[{\tt groupscale=25},][]{Kaastra_16}.
To fit the data, we adopted the range of 0.6-10 keV for the EPIC-pn spectrum.  

The obsid. 0111230101  was already analyzed by \cite{iaria_15};  the authors suggested the presence of an absorption line, discussed as cyclotron absorption line,  close to 0.7 keV by analyzing the EPIC-pn and RGS  spectra in the 0.6-10 keV  and 0.35-2 keV energy band, respectively. 
The observation was carried out   on 2001 March 7  for a duration of 54 ks. During the observation the EPIC-pn camera was operating in Timing Mode. We reduced the \textit{XMM-Newton} data using the Science Analysis Software (SAS) v21.0.0.

To mitigate the effects of particle flaring background, we selected the time intervals with a
rate lower than 1.5 c/s. 
We show the source and background EPIC-pn light curve in the 0.6-10 keV energy range cleaned from the particle flaring background in Fig. \autoref{fig:lightcurve_078482010} (right panel). The source light curve shows  the orbital modulation and ranges between 30  c s$^{-1}$ and 70 c s$^{-1}$, two partial eclipses are present close to  20000 s and 40000 s from the beginning of the observation.  

The EPIC-pn spectrum and background were extracted selecting  RAWX between 31 and 44 and RAWX between 3 and 5. The EPIC-pn has an exposure time of 41 ks, it was grouped  as shown above.

The two XMM observations are distant in time, we produced the hardness ratios (HRs) using the energy band 0.6-2 keV and 2-10 keV to check a possible change in the shape spectrum. The HRs are shown in Fig. \ref{fig:hr}. 
We observe that the HR value is  approximately 2 for the observation carried out  in 2001 and approximately 1.7 for the one carried out in 2017.  

Since the  NICER  spectrum alone may suggest that the addition of two edges at 0.56 and 0.71 keV does not require the presence of an additional cyclotron line at 0.7 keV although we note that the data-to-model ratio     below 1 keV is approximately 5\%, we decide to verify that this continues to be true when we perform a combined fit of the  NICER    and the two EPIC-pn spectra.   At 0.7 keV, the effective area of EPIC-pn is approximately 900 cm$^2$, which is only 30\% lower than that of  NICER, which is around 1250 cm$^2$. Given that the  NICER  spectrum has a systematic error larger than 1\%  and an exposure time of 11 ks, while both  EPIC-pn spectra  have  an exposure time of 41 ks, we can compare the statistics of the three spectra at 0.7 keV without the risk of biasing the fit toward the  NICER spectrum.
\begin{table*} 
\setlength{\tabcolsep}{3.5pt}
        \centering
        \scriptsize
        \caption{Best-fit parameters combining  PN1, PN2  and   NICER   spectra} 
                \label{tab:compare_spectrum}
        \begin{tabular}{ll | ccc |ccc| ccc}
        \toprule
                    &  & \multicolumn{3}{c|}{\texttt{Model I}}&
                    \multicolumn{3}{c|}{\texttt{Model L}} & \multicolumn{3}{c}{\texttt{Model M}}\\
        Components & Parameters & 
         NICER&  PN1   & PN2   
        &  NICER&  PN1   & PN2        
        & NICER&  PN1   & PN2 \\      
        \midrule
        {\tt Const}&   
        & [1] & $0.85\pm0.02$ &     $0.686\pm0.015$  
        & [1] &  $0.86\pm0.02$ & $0.688\pm0.015$ 
        & [1] &  $0.86\pm0.02$ & $0.685\pm0.015$ \\ 

       {\tt Edge} & E (keV) & 
         \multicolumn{3}{c|}{  [2.1]    } &     
         \multicolumn{3}{c|}{  [2.1]    }   &    
         \multicolumn{3}{c}{  [2.1]    }  \\

& $\tau$ &
$0.09  \pm0.02$  & [0] &[0]&  
$0.08  \pm0.02$  & [0] &[0]&
$0.08  \pm0.02$  & [0] &[0]\\    
 
   {\tt gabs} & E$_0$ (keV) & 
    \multicolumn{3}{c|}{$0.728\pm0.005$}      &
    \multicolumn{3}{c|}{$0.749^{+0.012}_{-0.027}$}  &
    \multicolumn{3}{c}{$0.744^{+0.011}_{-0.018}$}  \\

    & $\sigma_0 $ (keV) & 
    \multicolumn{3}{c|}{$0.097 \pm 0.008$}  &
    \multicolumn{3}{c|}{$0.075 \pm 0.013$} &
    \multicolumn{3}{c}{$0.081^{+0.016}_{-0.013}$}  \\

    &  Strength ($\times 10^{-2}$)&  
     \multicolumn{3}{c|}{  $5.9\pm0.9$}&
     \multicolumn{3}{c|}{  $3.3^{+1.3}_{-0.9}$}&
     \multicolumn{3}{c}{  $4.0^{+1.4}_{-1.0}$}\\

        {\tt TBabs}  & N$_{H}$& 
        \multicolumn{3}{c|}{$6.2\pm0.7$}&
        \multicolumn{3}{c|}{$6.4\pm0.7$} &
        \multicolumn{3}{c}{--} \\ 

        {\tt TBfeo}  & N$_{H}$& 
        \multicolumn{3}{c|}{--}&
        \multicolumn{3}{c|}{--} &
        \multicolumn{3}{c}{$6.3\pm0.7$} \\
 
           & O & 
        \multicolumn{3}{c|}{--}&
        \multicolumn{3}{c|}{--} &
        \multicolumn{3}{c}{$2.0\pm0.7$} \\

           & Fe & 
        \multicolumn{3}{c|}{--}&
        \multicolumn{3}{c|}{--} &
        \multicolumn{3}{c}{$<3.7$} \\

       {\tt Edge} & E (keV) & 
         \multicolumn{3}{c|}{ --   } &     
         \multicolumn{3}{c|}{  [0.56]    } &   
              \multicolumn{3}{c}{ --   } \\     
    
& $\tau$ &
 \multicolumn{3}{c|}{ --   } & 
  \multicolumn{3}{c|}{ $0.18\pm0.08$  } & 
   \multicolumn{3}{c}{ --   } \\  

       {\tt Edge} & E (keV) & 
         \multicolumn{3}{c|}{ --   } &     
         \multicolumn{3}{c|}{  [0.71]    } &   
              \multicolumn{3}{c}{ --   } \\     
    
& $\tau$ &
 \multicolumn{3}{c|}{ --   } & 
  \multicolumn{3}{c|}{ $<0.09$  } & 
   \multicolumn{3}{c}{ --   } \\

                 {\tt bbodyrad} &  kT$_{bb}$ (keV) & 
    \multicolumn{3}{c|}{ $0.182 \pm 0.004$  }    &
    \multicolumn{3}{c|}{ $0.185 \pm 0.004$  }    &
    \multicolumn{3}{c}{ $0.184 \pm 0.004$  }    \\

                      & N$_{bb}$ & 
              $2000^{+400}_{-300}$ &
              $3000^{+500}_{-400}$ & 
              ($2000^{+400}_{-300}$) &
              $2000\pm300$&
              $3000\pm400$ & 
              ($2000\pm300$) &
              $2000\pm300$ &
              $3000\pm400$ & 
              ($2000\pm300$) \\

         {\tt Comptb} &  kT$_0$ (keV) & 
          $0.90\pm 0.04$&
         $0.81\pm 0.04  $  &
         $0.80 \pm 0.03   $  &
        $0.91 \pm 0.04$&
         $0.82\pm 0.04  $  &
         $0.81 \pm 0.03   $  &
         $0.90 \pm 0.04$&
         $0.81\pm 0.04  $  &
         $0.80 \pm 0.03   $   \\
         
                      & kT$_e$ (keV)  & 
         $3.20^{+0.12}_{-0.10}$   &
         ($3.20^{+0.12}_{-0.10}$)   &
         $2.67\pm 0.05$ &
         $3.23^{+0.13}_{-0.10}$   &
         ($3.23^{+0.13}_{-0.10}$)   &
         $2.69\pm 0.05$     &
         $3.22^{+0.12}_{-0.10}$   &
         ($3.22^{+0.12}_{-0.10}$)   &
         $2.68\pm 0.05$          \\
               
                      &  $\alpha$ & 
                     $0.40 \pm 0.03 $ &
                     $0.32 \pm 0.02 $ &
                     $0.287 \pm 0.013  $ &
                     $0.41 \pm 0.03 $ &
                     $0.33 \pm 0.02 $ &
                     $0.291 \pm 0.015  $ &
                     $0.40 \pm 0.03 $ &
                     $0.32 \pm 0.02 $ &
                     $0.288 \pm 0.013  $ \\
                         
    & N$_{Comptb}$  ($\times 10^{-3}$)& 
 \multicolumn{3}{c|}{$4.4\pm0.2$}    &
 \multicolumn{3}{c|}{$4.4\pm0.2$}   &
 \multicolumn{3}{c}{$4.4\pm0.2$}     \\ 

 {\tt Gauss}    & E (keV)  
& \multicolumn{3}{c|}{[1.182]}     
& \multicolumn{3}{c|}{[1.182]}    
& \multicolumn{3}{c}{[1.182]}    \\ 

& $\sigma$ (eV) 
& \multicolumn{3}{c|}{[5] }     
& \multicolumn{3}{c|}{[5] }    
& \multicolumn{3}{c}{[5] }    \\

& I ($\times 10^{-5}$)  
& [0]&  $7\pm2 $  & $5\pm2 $   
& [0]&  $6\pm3 $  & $5\pm2 $    
& [0]&  $6\pm3 $  & $5\pm2 $    \\

 {\tt Gauss}    & E (keV)  
& \multicolumn{3}{c|}{[1.343]}    
& \multicolumn{3}{c|}{[1.343]}   
& \multicolumn{3}{c}{[1.343]}   \\ 

& $\sigma$ (eV) 
&\multicolumn{3}{c|}{[5]}   
&\multicolumn{3}{c|}{[5]}  
&\multicolumn{3}{c}{[5]}\\  

& I ($\times 10^{-5}$)  
& [0]&  $11\pm2 $     &  $5\pm2$       
& [0]&  $10\pm2 $     &  $4\pm2$      
& [0]&  $10\pm2 $     &  $4\pm2$       \\

 {\tt Gauss}    & E (keV)  
& \multicolumn{3}{c|}{[1.89]}    
& \multicolumn{3}{c|}{[1.89]}   
& \multicolumn{3}{c}{[1.89]}   \\ 

& $\sigma$ (eV) 
&\multicolumn{3}{c|}{[0]}   
&\multicolumn{3}{c|}{[0]}  
&\multicolumn{3}{c}{[0]}\\  

& I ($\times 10^{-5}$)  
& [0]&  $7\pm2 $     &  [0]      
& [0]&  $7\pm2 $     &  [0]        
& [0]&  $7\pm2 $     &   [0]         \\

 {\tt Gauss}    & E (keV)  
&\multicolumn{3}{c|}{$2.23\pm0.02$}    
&\multicolumn{3}{c|}{$2.23\pm0.02$}   
&\multicolumn{3}{c}{$2.23\pm0.02$}   \\

& $\sigma$ (eV)  
&\multicolumn{3}{c|}{[0]} 
&\multicolumn{3}{c|}{[0]} 
&\multicolumn{3}{c}{[0]}\\

& I ($\times 10^{-4}$)  
 & [0] &[0]   &$-0.9\pm 0.2$    
 & [0] &[0]   &$-0.9\pm 0.2$    
 & [0] &[0]   &$-0.9\pm 0.2$    \\

 {\tt Gauss}    & E (keV)  
& \multicolumn{3}{c|}{$6.31\pm0.08$}      
& \multicolumn{3}{c|}{$6.31\pm0.08$}     
& \multicolumn{3}{c}{$6.31\pm0.08$}     \\ 

& $\sigma$ (eV) 
& \multicolumn{3}{c|}{$440\pm70$ }      
& \multicolumn{3}{c|}{$440\pm70$ }     
& \multicolumn{3}{c}{$440\pm70$ }     \\

& I ($\times 10^{-4}$)  
& \multicolumn{3}{c|}{$3.8\pm0.7 $ }    
& \multicolumn{3}{c|}{$3.9\pm0.8 $ }    
& \multicolumn{3}{c}{$3.9\pm0.8 $ } \\

{\tt Gauss}  & E (keV)  
&   $6.412 ^{+0.008}_{-0.010}$ 
&   $6.505 \pm 0.008  $    
& ($6.412 ^{+0.008}_{-0.010}$) 
&   $6.412 ^{+0.007}_{-0.010}$ 
&   $6.505 \pm 0.008  $    
& ($6.412 ^{+0.007}_{-0.010}$) 
&   $6.412 ^{+0.008}_{-0.010}$ 
&   $6.505 \pm 0.008  $    
& ($6.412 ^{+0.008}_{-0.010}$)\\

        \ion{Fe}{i}  & $\sigma$ (eV) 
& \multicolumn{3}{c|}{ [70] }     
& \multicolumn{3}{c|}{ [70] }     
& \multicolumn{3}{c}{ [70] }    \\

& I ($\times 10^{-4}$)  
&   $3.5\pm0.3 $  &  $5.1\pm0.4 $ & ($3.5\pm0.3 $)  
&   $3.5\pm0.3 $  &  $5.0\pm0.4 $ & ($3.5\pm0.3 $)  
&   $3.5\pm0.3 $  &  $5.1\pm0.4 $ & ($3.5\pm0.3 $)\\  
  
 {\tt Gauss}     & E (keV) 
&   $6.954^{+0.021}_{-0.013} $   
&   $7.056 \pm 0.015$   
&   ($6.954^{+0.021}_{-0.013} $)    
&   $6.954^{+0.021}_{-0.013}$   
&   $7.054 ^{+0.015}_{-0.007}$   
&   ($6.954^{+0.021}_{-0.013}$)   
&   $6.955^{+0.020}_{-0.013}$   
&   $7.054 \pm 0.015$   
&   ($6.955^{+0.020}_{-0.013}$)   \\

        \ion{Fe}{xxvi} & $\sigma$ (eV) 
& \multicolumn{3}{c|}{ [70]  }   
& \multicolumn{3}{c|}{ [70]  }    
& \multicolumn{3}{c}{ [70]  }   \\

& I ($\times 10^{-4}$) 
&  $1.9 \pm 0.3  $ &  $2.9 \pm 0.3  $& ($1.9 \pm 0.3$) & $1.9 \pm 0.3  $ &  $2.9 \pm 0.3  $& ($1.9 \pm 0.3$)  
& $1.9 \pm 0.3  $ &  $2.9 \pm 0.3  $& ($1.9 \pm 0.3$) \\

          &  $\chi^2$/d.o.f.  
          & \multicolumn{3}{c|}{494/356}  
          & \multicolumn{3}{c|}{484/354}    
          & \multicolumn{3}{c}{489/354 }  \\ 
 \bottomrule
\end{tabular}
    \begin{tablenotes}
\item[]  The errors are at 90\% confidence level. The values in square brackets weare kept  fixed during the fit. The values in round brackets were constrained to the values of the  NICER  spectrum.  The $N_H$ value is in units of $ 10^{20}$  cm$^{-2}$. 
    \end{tablenotes}
    \end{table*} 

Initially, we fitted  the  NICER   spectrum (0.3-10 keV),  the 0.6-10 keV EPIC-pn spectrum corresponding to obsid. 0784820101 (hereafter PN1 spectrum), the 0.6-10 keV EPIC-pn spectrum corresponding to obsid. 0111230101 (hereafter PN2 spectrum) using \texttt{Model B}, to which we added  a Gaussian component in the Fe-K region of the spectrum because   PN1 and PN2  spectra require it. We added  a constant to accommodate the different normalization between the three spectra. The constant value was held fixed at 1 for the  NICER   spectrum, while it was allowed to vary for PN1 and PN2 spectra. To take into account the systematic feature  present close 2 keV in the   NICER spectrum   (see \autoref{fig:modelGsyscal}) we added an absorption edge   with energy threshold fixed at 2.1 keV; we kept fixed to zero the depth of the edge for the other spectra.
 In addition, since the three spectra were extracted from non-simultaneous observations, we   allowed 
 the normalization of the blackbody component, the seed-photon temperature and 
 the energy index $\alpha$ of the  \texttt{Comptb}  component to vary independently.  
 We forced the value of the electronic temperature of the PN1 spectrum to that of the NICER spectrum. 
 We   left to vary independently the energies and the normalizations of the two narrow Gaussian line at 6.4 and 6.96 keV for PN1 spectrum and kept fixed  at 70 eV their widths. Finally, we   added two Gaussian emission lines at PN1 and PN2 spectrum at 1.18 keV and 1.343 keV with width fixed at 5 eV. 
 To take into account of the systematic at 2 keV in the PN2 spectrum we added a Gaussian component with null width and negative normalization (hereafter we call this model \texttt{Model H}). 

 By fitting the spectrum, we obtained an unacceptable fit with a $\chi^2$(dof) of 1337(359), the ratio below 1 keV shows that the model deviate of 20\% from the data (top-left panel of \autoref{fig:nicer_epn07}).   Then, we added to model a \texttt{gabs} component at 0.7 keV (hereafter \texttt{Model I}). The best-fit improves significantly; we found a $\chi^2$(dof) of 494(356) with a $\Delta \chi^2=843$. The ratio is within 2\% as shown  in the bottom-left panel  of \autoref{fig:nicer_epn07}.
 The best-fit parameters are shown in the third column of \autoref{tab:compare_spectrum}.  The best-fit parameters of the \texttt{gabs} component are $E_0=0.728 \pm 0.005$ keV, $\sigma_0=0.097\pm 0.008$ keV and Strength$=0.059\pm 0.009$, these results are consistent with those obtained by fitting the  NICER    spectrum alone. 
 
 To test the significance of adding the \texttt{gabs} component, we simulated 1000 spectra using the Monte-Carlo method as described in the previous section. The histogram of the $\Delta \chi^2$ values obtained from the simulation is shown in the left panel of \autoref{fig:histo_ni_pn1_pn2} and it is well-fitted by a $ \chi^2$ with three d.o.f. (red curve). We found that the largest value of $ \Delta \chi^2$ obtained from the simulation is 15.9, the probability to observe a $ \Delta \chi^2$ of 849 by chance  is $5.2 \times 10^{-180}$ corresponding to a statistical  significance of  29 $\sigma$.

We started from \texttt{Model H} adding to it two absorption edges at 0.56 keV and 0.71 keV. By fitting the three spectra we found  a $\chi^2$(dof) of 
615(357),   the data deviate from the model up to 5\% below 0.6 keV as shown in the top-center panel of \autoref{fig:nicer_epn07}.
We added to the model a \texttt{gabs} component at 0.7 keV (hereafter \texttt{Model L}). The best-fit improves significantly; we found a $\chi^2$(dof) of 484(354) with a $\Delta \chi^2=131$. We show the   residuals in the bottom-center   panel of \autoref{fig:nicer_epn07}. The best-fit parameter are shown in the fourth  column of \autoref{tab:compare_spectrum}. The best-fit parameters of the \texttt{gabs} component are $E_0=0.749^{+0.012}_{-0.027}$ keV, $\sigma_0=0.075\pm0.013$ keV and Strength$=0.033^{+0.013}_{-0.009}$, these results are consistent with those obtained by adopting \texttt{Model I}. 
 
Also for  this model, we  assessed the significance of adding the \texttt{gabs}    component as done previously. The histogram of the simulated spectra is shown in the middle panel of \autoref{fig:histo_ni_pn1_pn2}. 
We found that the largest value of $ \Delta \chi^2$ obtained from the simulation is 16.7, the probability to observe a $ \Delta \chi^2$ of 131 by chance is $8.6 \times 10^{-26}$ corresponding to a statistical  significance of  10.5 $\sigma$.  
 Our test indicates that the addition of a component representing a cyclotron line at 0.7 keV is necessary, even though two edges are included in the model  at 0.56 and 0.71 keV.

Finally, instead of adding two edges at 0.56 and 0.71 keV  we used a self-consistent absorption model. Specifically, we replaced \texttt{TBabs} with \texttt{TBfeo}, allowing the abundances of oxygen and iron to vary freely,   as suggested by the  NICER  team.
By fitting the  spectra we found  a $\chi^2$(dof) of 658(357); also in this case  the data deviate from the model up to 5\% below 0.6 keV  (top-right panel of \autoref{fig:nicer_epn07}).  
We added to the model a \texttt{gabs} component at 0.7 keV (hereafter \texttt{Model M}). We found a $\chi^2$(dof) of 489(354) with a $\Delta \chi^2=169$.  The best-fit parameters are shown in the fifth  column of \autoref{tab:compare_spectrum}, the residuals in the bottom-right panel of \autoref{fig:nicer_epn07}).  The best-fit parameters of the \texttt{gabs} component are $E_0=0.744^{+0.011}_{-0.018}$ keV, $\sigma_0=0.081^{+0.016}_{-0.013}$ keV and Strength$=0.040^{+0.014}_{-0.010}$, these results are consistent with those obtained by adopting \texttt{Model I} and \texttt{Model L}. 
In this case,
we find that the largest value of $ \Delta \chi^2$ obtained from the simulation is 17.2, the probability to observe a $ \Delta \chi^2$ of 169 by chance  is $5.2 \times 10^{-34}$ corresponding to a statistical  significance of  12 $\sigma$  
 (right panel of \autoref{fig:histo_ni_pn1_pn2}). 
\begin{table*}
        \centering
        \caption{Best-fit parameters combining    NICER  and \textit{NuSTAR} spectra} 
                \label{tab:spectrum}
        \scriptsize  
        \begin{tabular}{ll   |cc| cc |cc}
        \toprule
        Components & Parameters &
 \multicolumn{2}{c|}{Model 1}  &
\multicolumn{2}{c|}{Model 2}  &
\multicolumn{2}{c}{Model 3}  \\
         &  & NICER  & \textit{NuSTAR}&  NICER  & \textit{NuSTAR}&  NICER  & \textit{NuSTAR}
     \\  
        \midrule 

        {\tt Const$_{\tt instr}$}&   &  
        [1]   & $1.192\pm0.011$& 
        [1] & $1.170\pm0.014$& 
        [1]  & $1.190\pm0.013$\\ 
        
       {\tt Edge} & E (keV) & 
         \multicolumn{2}{c|}{$ [2.1] $  } &\multicolumn{2}{c|}{$ [2.1] $  } &\multicolumn{2}{c}{$ [2.1] $  } \\    
            & $\tau$ &
          $0.07  \pm0.02$  & 
          [0] &$0.05\pm0.02$ &[0]&$0.04\pm0.02$ &[0]\\

                      {\tt Edge} & E (keV) & 

        \multicolumn{2}{c|}{$9.60\pm 0.08 $} 
    &\multicolumn{2}{c|}{$9.62\pm0.09$} &\multicolumn{2}{c}{$9.62\pm0.09$}\\
            & $\tau$ &
         [0]& $0.047 \pm 0.006 $ &[0]&$0.038\pm0.007$&[0]&$0.037\pm0.007$\\           
 
        {\tt TBabs}  & N$_{H}$ ($\times 10^{20}$  cm$^{-2}$)& 
          \multicolumn{2}{c|}{$6.3\pm1.0$} &\multicolumn{2}{c|}{$6.6\pm1.0$}
&\multicolumn{2}{c}{$8\pm2$}\\

   {\tt gabs} & E$_0$ (keV) & 
     \multicolumn{2}{c|}{$0.730 \pm 0.013 $} &\multicolumn{2}{c|}{$0.724 \pm 0.014 $} 
&\multicolumn{2}{c}{$0.66 \pm 0.03 $}\\
                  & $\sigma_0 $ (keV) & 
                    \multicolumn{2}{c|}{$0.11 \pm 0.02$}&
       \multicolumn{2}{c|}{$0.10\pm0.02$} &
       \multicolumn{2}{c}{$0.17\pm0.03$} \\
                  & $\tau_0$  &  
                 \multicolumn{2}{c|}{$0.25 \pm 0.10$}&
                 \multicolumn{2}{c|}{$0.25 \pm 0.10$} &
                 \multicolumn{2}{c}{$0.4 \pm 0.2$}\\

                 {\tt bbodyrad} &  kT$_{bb}$ (keV) & 
         \multicolumn{2}{c|}{$0.187\pm0.006$}  &
         \multicolumn{2}{c|}{$0.173\pm0.007$} &\multicolumn{2}{c}{$0.158\pm0.011$}  \\
                      & N$_{bb}$  & 
                  \multicolumn{2}{c|}{$2300^{+500}_{-400}$} &  
    \multicolumn{2}{c|}{$2800^{+800}_{-500}$} &   
    \multicolumn{2}{c}{$5000^{+2000}_{-1000}$}    \\

         {\tt Comptb} &  kT$_0$ (keV) & 
         \multicolumn{2}{c|}{$1.08 \pm 0.03 $} &          \multicolumn{2}{c|}{$1.34 \pm 0.05 $}   
&          \multicolumn{2}{c}{$1.32^{+0.03}_{-0.05}  $}   \\
                                & kT$_e$ (keV)  & 
                   \multicolumn{2}{c|}{$4.63 \pm 0.02  $}  &
 \multicolumn{2}{c|}{$4.22 \pm 0.07$}    &
 \multicolumn{2}{c}{$4.34\pm0.07$}   \\                      &  $\alpha$ & 
                      $0.32 \pm 0.02\;$ &
                      $0.459\pm0.011$&
                      $<0.13$& 
                      $0.22\pm0.06$&
                      $0.19^{+0.05}_{-0.07}$& 
                      $0.35\pm0.07$\\
                      
                       &  log$A$ & 
                      \multicolumn{2}{c|}{$0.77 \pm 0.06 $}  &
    \multicolumn{2}{c|}{$0.20 \pm 0.10 $}   &
    \multicolumn{2}{c}{$0.35 \pm 0.12 $}  \\
                                          
                      & N$_{Comptb}$  ($\times 10^{-3}$)& 
                       \multicolumn{2}{c|}{$6.0\pm1.3 $} &
\multicolumn{2}{c|}{$6.55\pm0.15  $} &
\multicolumn{2}{c}{$6.00^{+0.17}_{-0.11} $} 
                       \\ 

         {\tt rdblur} &  Betor   & 
         \multicolumn{2}{c|}{--} &
         \multicolumn{2}{c|}{--}  &
\multicolumn{2}{c}{$-3.1^{+0.5}_{-1.7}  $} \\
                       &  R$_{in}$ (R$_g$)& 
                      \multicolumn{2}{c|}{--} &
\multicolumn{2}{c|}{--}  &
\multicolumn{2}{c}{$62^{+21}_{-13}  $  }   \\
                       &  R$_{out}$ (R$_g$)& 
                      \multicolumn{2}{c|}{--}  &
  \multicolumn{2}{c|}{--} &                     \multicolumn{2}{c}{[3000]   }   \\
                       &  $i$ (deg)& 
                      \multicolumn{2}{c|}{--}  & \multicolumn{2}{c|}{--} &
                      \multicolumn{2}{c}{[82.5]}   \\ 

         {\tt rfxconv} &  f$_{refl}$   & 
         \multicolumn{2}{c|}{--}   &
 \multicolumn{2}{c|}{--} &        \multicolumn{2}{c}{$0.66^{+0.08}_{-0.10} $}   \\
                       &  log$(\xi)$ & 
                      \multicolumn{2}{c|}{--}   &
 \multicolumn{2}{c|}{--} &                      \multicolumn{2}{c}{ $1.4 ^{+0.4}_{-1.4}$ }   \\ 

 {\tt expabs} &   $E_c $ (keV)   &
\multicolumn{2}{c|}{--}&
\multicolumn{2}{c|}{ [kT$_0$]} &
\multicolumn{2}{c}{ [kT$_0$]} \\

  {\tt power-law} &   $\Gamma $   & 
         \multicolumn{2}{c|}{--}   &
         \multicolumn{2}{c|}{$1.96\pm0.05$}    &      \multicolumn{2}{c}{$1.99\pm0.05$}   \\
                       &  N$_{po}$  ($\times 10^{-2}$) & 
            \multicolumn{2}{c|}{--}   &
            \multicolumn{2}{c|}{ $2.1\pm 0.5 $  }  &
            \multicolumn{2}{c}{ $2.2\pm 0.4 $  }  \\ 

       {\tt Gauss}    & E (keV)  
         & \multicolumn{2}{c|}{$6.12^{+0.08}_{-0.10}$}    & 
\multicolumn{2}{c|}{ $6.19^{+0.10}_{-0.34}$  }         & \multicolumn{2}{c}{ -- }    \\
        & $\sigma$ (eV) 
        & \multicolumn{2}{c|}{$740^{+110}_{-90}$ }    
&\multicolumn{2}{c|}{ $530^{+90}_{-250}$  }   & \multicolumn{2}{c}{-- }    \\
                        & I ($\times 10^{-4}$)  
& \multicolumn{2}{c|}{$9.9^{+1.6}_{-1.3} $ }   &  \multicolumn{2}{c|}{ $6.1^{+1.5}_{-3.4}  $}
& \multicolumn{2}{c}{ -- }   \\

        {\tt Gauss}  & E (keV)  
         & \multicolumn{2}{c|}{ $6.378 \pm 0.014  $  }    
          & \multicolumn{2}{c|}{ $6.39 \pm 0.02    $  }     
        & \multicolumn{2}{c}{ $6.371^{+0.035}_{-0.012}      $  }    \\
        \ion{Fe}{i}  & $\sigma$ (eV) 
        & \multicolumn{2}{c|}{ $80^{+20}_{-30}$  }   
        & \multicolumn{2}{c|}{ $70^{+40}_{-30}      $   }
        & \multicolumn{2}{c}{ $30^{+40}_{-20}      $   }\\ 
                        & I ($\times 10^{-4}$)  

& \multicolumn{2}{c|}{ $3.3\pm0.5 $  }   
& \multicolumn{2}{c|}{ $3.2^{+0.8}_{-1.6}$  } 
& \multicolumn{2}{c}{ $2.5 \pm0.7 $  }   \\ 
 
       {\tt Gauss}     & E (keV) 
& \multicolumn{2}{c|}{ $6.90 \pm 0.03  $  }    & \multicolumn{2}{c|}{ $6.91 \pm 0.03  $  } 
& \multicolumn{2}{c}{ $6.95 \pm 0.06  $  }    \\
        \ion{Fe}{xxvi} & $\sigma$ (eV) 
& \multicolumn{2}{c|}{ $80^{+20}_{-30}$  }  
& \multicolumn{2}{c|}{$70^{+40}_{-30}$  } 
& \multicolumn{2}{c}{ $30^{+40}_{-20} $  }\\ 
                        & I ($\times 10^{-4}$) 
&\multicolumn{2}{c|}{ $1.7 \pm 0.3  $ }
&\multicolumn{2}{c|}{ $1.8 ^{+1.0}_{-1.5} $  } &\multicolumn{2}{c}{ $0.7 \pm 0.2  $  }   \\

          &  $\chi^2$/d.o.f.  
          & \multicolumn{2}{c|}{469/318}  
          & \multicolumn{2}{c|}{302/316}  
          & \multicolumn{2}{c}{296/315}\\

\bottomrule
\end{tabular}
    \begin{tablenotes}
\item[] The errors are at 90\% confidence level. The values in square brackets were kept  fixed during the fit. The intensity of the lines is in units of  photons/cm$^2$/s. 
    \end{tablenotes}
\end{table*}
Since the component associated with a 0.7 keV cyclotron line is required for all three described models,   we discuss a model that requires a cyclotron line at low energies, with abundances of O and Fe fixed to solar values (\texttt{Model G}) in the subsequent sections of the work.

However, since the reduced-$\chi^2$  associated with the  NICER  spectrum  alone is much smaller than one and attributable to an overestimation of systematic errors in the 0.3-10 keV range (see the third column in \autoref{Tab:model_G}), we  extracted the spectrum by assigning a 1\% systematic error to our data. 
Fitting the  NICER  spectrum adopting \texttt{Model G}, we obtained a $\chi^2$   of 117(138) with the ratio between 0.3 and 10 keV   below 2\%. The best-fit values  shown in the fourth column of \autoref{Tab:model_G} indicate that the spectral shape  remains unchanged when transitioning from the systematic error suggested by the  NICER  calibration team to a 1\% systematic error, while the reduced-$\chi^2$ increases from 0.61 to 0.87.

\subsection{Combined   NICER-\textit{NuSTAR} spectra}
 \begin{figure*}
    \centering
    \includegraphics[scale=0.55]{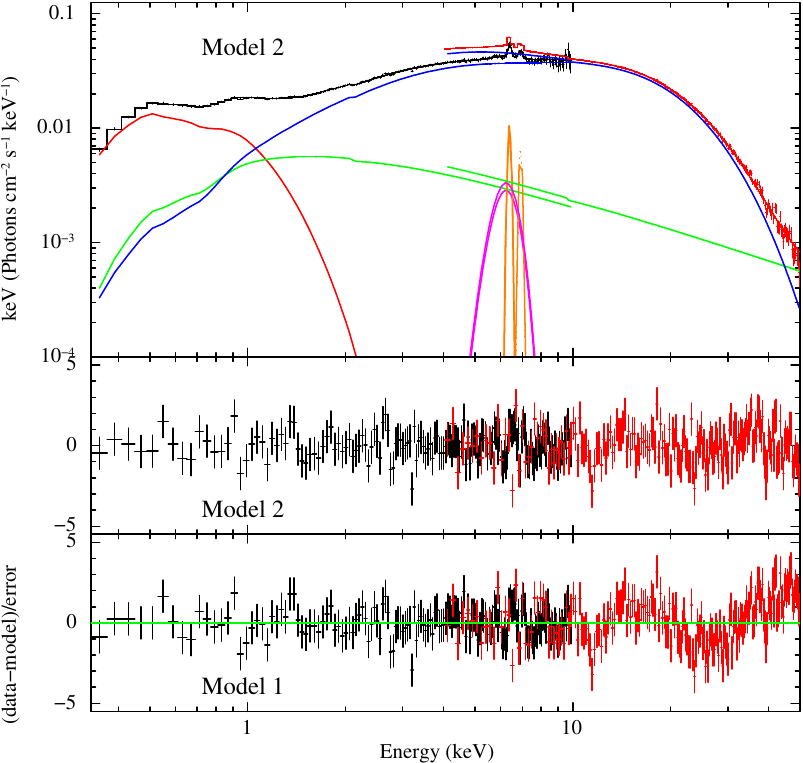} \hspace{1cm}
    \includegraphics[scale=0.55]{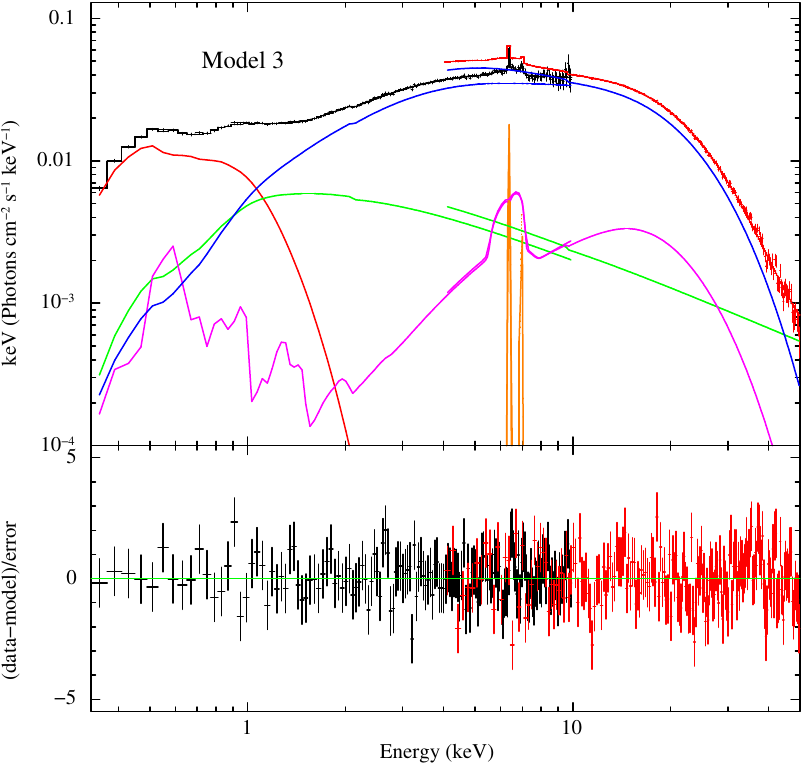}
    \caption{Unfolded spectrum of {\tt Model 2}  (top-left panel) using  a  0.3-10 keV  NICER/XTI spectrum  (black color) combined with a 4-50 keV  
    \textit{NuSTAR}/(FPMA+FPMB)   spectrum (red color), the colors of the components are described in the text. Residuals adopting {\tt Model 2}   and adopting {\tt Model 1}   are shown in the middle-left panel and bottom-left panel, respectively. Unfolded spectrum and
    residuals of {\tt Model 3} in top-right and bottom-right panel, respectively.}
    \label{fig:spectrum}
\end{figure*} 
We fitted  the combined 0.3-10 keV   NICER spectrum and the 4-50 keV {\it NuSTAR} spectrum using    \texttt{Model G}   where we left free to vary the parameter $logA$ of the \texttt{Comptb} component. To take into account the systematic feature  present close 2 keV in the   NICER  spectrum   (see \autoref{fig:modelGsyscal}) we added an absorption edge  with energy threshold fixed at 2.1 keV; we kept fixed to zero the depth of the edge for the {\it NuSTAR} spectrum. In addition, since the two spectra are extracted from non-simultaneous observations, we   allowed the energy index $\alpha$ of the  \texttt{Comptb}  component to vary independently between the two spectra. We added a constant to account for the different normalization between the two spectra. The constant value was kept fixed at 1 for the   NICER   spectrum, while it was free to vary for the {\it NuSTAR}   spectrum. Finally, we kept fixed  at 70 eV the widths of the two lines in the Fe-K region.

Fitting the spectrum, we obtained an unacceptable fit with a $\chi^2$(dof) of 1038(324). In the {\it NuSTAR}    spectrum, a feature at 9.6 keV is evident, which we modeled with an absorption edge. We set the depth of the edge to zero for the   NICER     spectrum.
 The addition of the absorption edge improves the fit; we obtained a $\chi^2$(dof)  of 905(322). However, large residuals were still present in the Fe-K region of the spectrum. For this reason, we added an additional Gaussian component to the model.
The model (hereafter {\tt Model\;1}) is: 
\begin{equation*}
\begin{split} 
 {\tt Model\;1:  const*gabs*edge*edge*TBabs*(bbodyrad+}\\
{\tt +Comptb+gauss+gauss+gauss). }
\end{split}
\end{equation*}

We found a $\chi^2$(d.o.f.) value of 469(319), the added  Gaussian emission line has a centroid of   
$6.13\pm0.08$ keV and a $\sigma$ of $730\pm90$ eV. 
To better constrain this broad line, we left the width  of one of the two narrow lines free to vary and constrained  the amplitude of the second narrow line to vary with the first one. The obtained fit was statistically equivalent; we found a $\chi^2$(dof) of 469(318). The best-fit parameters are reported in the third column of \autoref{tab:spectrum}, the corresponding residuals are shown in the bottom-left panel of \autoref{fig:spectrum}. Analyzing the residuals, it is evident that the \textit{NuSTAR} spectrum (in red) is not well modeled above 15 keV.  We added a power-law component with a cut-off at low energies to fit the residuals at high energies. We fixed the energy of the cut-off component ({\tt expabs} in XSPEC) to the seed-photon temperature of the {\tt Comptb} component.
The model (hereafter {\tt Model\;2}) is: 
\begin{equation*}
\begin{split} 
 {\tt Model\;2:  const*gabs*edge*edge*TBabs*(bbodyrad+}\\
{\tt +expabs*po+Comptb+gauss+gauss+gauss). }
\end{split}
\end{equation*} 
where {\tt po} indicates the power-law component.

By adding the absorbed  power-law component we obtained a $\chi^2$(d.o.f.) of 302(316) and the residuals above 15 keV disappeared. We show the best-fit parameters in  
the fourth column of \autoref{tab:spectrum}, the unfolded spectrum and the residuals are show in the top-left and middle-left panel of \autoref{fig:spectrum}. 

In the unfolded spectrum 
we show the blackbody, the Comptonized component and the power-law in red, blue and green color, respectively. The narrow emission lines are in orange color and, finally, the broad Gaussian line is shown in magenta.   

The temperature of the thermal emission is $0.173 \pm 0.007$ keV,  the Comptonized component has a seed-photon temperature  of $1.34 \pm 0.05$ keV, an  electron temperature of $4.22\pm0.07$ keV, 
a $\alpha$-index $<0.13$ and  $0.22\pm0.06$ for   NICER  and {\it NuSTAR} spectrum respectively. 
The log$A$ parameter is  $0.20\pm 0.10$ implying that  ($39\pm6$)\% of the seed photons is directly seen by the observer while the remaining part is scattered in the Comptonizing cloud. The narrow emission lines are at $6.39\pm0.02$ keV and $6.91\pm0.03$ keV with a width of $70^{+40}_{-30}$ eV. The first line is associated with an emission line of neutral iron, the second line is marginally compatible with the presence of  \ion{Fe}{xxvi} ions. 

 The broad emission line has 
 an energy of $6.19^{+0.10}_{-0.34}$ keV and  a width of $530^{+90}_{-250}$ eV.  
  The photon-index of the power-law component  is $1.96\pm0.05$. 
 The {\tt gabs} component at low energies, that is the cyclotron line,  has a centroid at  
 $0.724\pm 0.014$ keV, a width of $0.10\pm0.02$ keV and an optical depth of   $0.25 \pm 0.10$. 
 Finally, an absorption edge at $9.62 \pm 0.09$ keV is observed.

 The unabsorbed fluxes in the 0.1-100 keV energy range are 
 $1.3 \times 10^{-9}$ erg s$^{-1}$ cm$^{-2}$, $2.6 \times 10^{-11}$ erg s$^{-1}$ cm$^{-2}$ and $1.4 \times 10^{-10}$ erg s$^{-1}$ cm$^{-2}$ associated with the Comptonized component, the blackbody and the power-law component, respectively. 
Assuming a distance to the source of 2.5 kpc, the total unabsorbed luminosity in the 0.1-100 keV energy range is 
$1.1 \times 10^{36}$ erg s$^{-1}$  {($6.6 \times 10^{36}$ erg s$^{-1}$ for  a distance to the source of 6.1 kpc, recently estimated by \cite{Arnason21}. Hereafter,  we assume that the distance to the source is 6.1 kpc.}
 
Following a scenario describing the presence of cyclotron line at 0.7 keV we changed the  {\tt gabs} component  with 
a  {\tt cyclabs} component. We obtained a statistically equivalent fit with a $\chi^2$(d.o.f.)$= 304(316)$; the best-fit values associated with the {\tt cyclabs} component are:
$E = 0.69 \pm 0.02$ keV, width$=0.12\pm0.03 $ keV and depth$= 0.31\pm0.04$.   These values are compatible with those reported by 
\cite{iaria_15}.

Because  we are observing a  broad Gaussian line in the Fe-K region  of the spectrum  we explored a scenario in which a reflection component from the accretion disk is present  as suggested by \cite{Anitra_21}.  To model the spectrum we changed the broad 
Gaussian line with a reflection component blurred by relativistic effects. We assumed that the incident radiation   comes from the Compton cloud. The model (hereafter {\tt Model 3}) is:
\begin{equation*}
\begin{split} 
{\tt Model\;3:  const*gabs*edge*edge*TBabs*(bbodyrad+}\\
{\tt +Comptb+ expabs*po+gauss+gauss+}\\
{\tt +rdlur*rfxconv*Comptb)},
\end{split}
\end{equation*} 
where {\tt rfxconv}  is the component that takes into account  the reflection component from the accretion disk while   {\tt rdblur} is the component that 
takes into account  the relativistic blurring.   We fixed the outer radius $R_{out}$ at 3000 gravitational radii after checking that the $\chi^2$ value was not sensible to  this parameter.   The inclination angle was fixed at $82.5^{\circ}$ 
as reported in literature \citep{mason_82}.  
 
 We show the best-fit values in the fifth column of  \autoref{tab:spectrum}, the unfolded spectrum and the residuals are shown in the top-right and bottom-right panels  of \autoref{fig:spectrum}. 
 In the unfolded spectrum of  {\tt Model 3}  the red, blue, magenta   and green component are the blackbody, the Comptonized component, the reflection component and the power-law. The two narrow emission lines in the Fe-K region are in orange. 

We obtained that $\chi^2$(d.o.f.)$= 296(315)$.   The reflection component has a reflection amplitude of  $0.66^{+0.08}_{-0.10}$, an inner disk radius of $62^{+21}_{-13}$ gravitation radii and  a ionization of the disk of log($\xi$)$=1.4^{+0.4}_{-1.4}$.   The best-fit values of the {\tt gabs} component are $E_0=0.66 \pm 0.03 $ keV,
$\sigma_0=0.17\pm0.03$ keV and an optical depth  $\tau_0=0.4 \pm 0.2$.  

The unabsorbed fluxes in the 0.1-100 keV energy range are 
 $1.2 \times 10^{-9}$ erg s$^{-1}$ cm$^{-2}$, $3.3 \times 10^{-11}$ erg s$^{-1}$ cm$^{-2}$,  $1.4 \times 10^{-10}$ erg s$^{-1}$ cm$^{-2}$ and  $1.2 \times 10^{-10}$ erg s$^{-1}$ cm$^{-2}$  associated with the Compton cloud, the blackbody component, the power-law component and the reflection component, respectively. 
The total unabsorbed luminosity in the 0.1-100 keV energy range is   $6.6 \times 10^{36}$ erg s$^{-1}$ for  a distance to the source of 6.1 kpc.

\section{Discussion}
Using the {\it NuSTAR} data  (obsid. 30701025002), we   updated the orbital ephemeris of X 1822-371 by adding one eclipse arrival time.  
 We find a $\dot{P}_{\rm orb}$ of $1.426(26) \times 10^{-10}$ s s$^{-1}$, compatible with the values present in literature \citep[see, e.g.,][]{Anitra_21}. As discussed before, the addition of a cubic term or a sinusoidal term to the quadratic model does not improve significantly the fit. 

We   estimated the spin period of the source during the 
two {\it NuSTAR} observations discussed above and confirmed the 
 presence of a linear long-term decrease of the spin period  
 finding $\dot{P}_{\rm spin}$ of $-2.64(2) \times 10^{-12}$ s s$^{-1}$,   compatible with the value reported by \cite{Bak17}. 

The analyzed energy spectrum was obtained by combining  the  
0.3-10 keV  NICER  spectrum  (obsid. 5202780101) with the  4.5-50 keV  \textit{NuSTAR}  spectrum.
The continuum emission is composed of a blackbody component plus a Comptonized component and a power-law component. 
The blackbody component has a temperature of $0.158\pm0.011$ keV; by  assuming a
distance to the source of 6.1 kpc  we infer an emission radius of $43 ^{+8}_{-5}$ km.  The seed-photon temperature and the electron temperature of the Comptonized component are  
kT$_0=1.32 ^{+0.03}_{-0.05} $ keV and kT$_e=4.34\pm0.07$ keV, respectively. The illumination factor  $logA$ is
$0.35 \pm 0.12$   and the $\alpha $ value, the energy index of the Comptonization spectrum,  is $0.19 ^{+0.05}_{-0.07}$ and $0.35 \pm 0.07$ for  NICER   and  \textit{NuSTAR} spectrum, respectively.   We find that the percentage of 
 seed-photon radiation directly seen by the observer  is $31\pm6$\% suggesting that the Comptonizing corona is   geometrically compact and does not completely surround the inner regions of the system.  The seed-photon emission radius is  $2.4 \pm 0.2$ km. 
 
 The {\tt gabs} component is interpreted as cyclotron  absorption line \citep[see also][]{iaria_15}, with a centroid energy of $E_0=0.66\pm0.03$ keV.  
The relation between the energy  of the fundamental harmonic and the magnetic field strength   is  $B_{12}=E_0/(11.6)\;(1-0.295\, m/R_6)^{-1/2}$ G, where $m$ is the NS mass in unit of solar masses and $R_6$ is the NS radius in units of $10^6$ cm. The NS mass was estimated to be between $1.61\pm0.15$ and $1.70\pm0.13$ M$_{\odot}$ \citep{iaria_15}. Assuming a NS radius of 10 km and the value of $E_0$,   we estimate that the magnetic field strength  is $B=(7.9\pm0.5) \times 10^{10}$ G and $B=(8.1\pm0.5) \times 10^{10}$ for a NS mass  of   $1.61\pm0.15$ M$_{\odot}$ and $1.70\pm0.13$ M$_{\odot}$, respectively. These values of the magnetic field strength are compatible with the estimate done by \cite{jonker_01} who calculated a value of $B \sim 8 \times 10^{10}$ G for a luminosity of the source of $10^{38}$
erg s$^{-1}$. In the following we discuss our results  for a NS mass of $1.61$ M$_{\odot}$ (similar results are obtained for 
a NS mass  of $1.70$ M$_{\odot}$).

{We observe the weakest magnetic field strength ever inferred from an electron cyclotron line. Since the formation of a cyclotron line requires that the magnetic field is $B >(v/c)^2 B_{\rm crit}$, where $v/c$ is the velocity of electrons (in units of the speed of light) perpendicular to the field lines and  $B_{\rm crit}= 4.4 \times 10^{13}$ G, we expect that the  electron velocity component perpendicular to the field lines is of the order of 10\% the speed of light to obtain a threshold value of $B$ of the order of $8 \times 10^{10}$ G.
}

To estimate the expected luminosity from the source 
 we used the \cite{ghosh_79} equation, which links the spin-period derivative, its magnetic field and its luminosity: 
\begin{equation}
-\dot{P}=5.0 \times 10^{-5} \; \mu_{30}^{2 / 7}\; n\left(\omega_{s}\right)\;m^{-3/7}\; R_6^{6/7}I_{45}^{-1}\left(P L_{37}^{3 / 7}\right)^{2} \mathrm{s}/ \mathrm{yr},
\end{equation}
where $I_{45}$ is the NS moment of inertia in units of 10$^{45}$ g cm$^{2}$,    $R_6$ is the NS radius in units of $10^6$ cm, $m$ is the NS mass in units of solar masses  and $P$ is the spin period of the source, that is, 0.591124781(13) s. The parameter n($\omega_{s}$) is the dimensionless accretion torque that is a function of the fastness parameter $\omega_{s}$. When $\omega_{s}$ < 0.95, we can use the following approximate expressions  \citep{ghosh_79}:
\begin{align}
\begin{split}
&n\approx 1.39\left\{1-\omega_{s}\left[4.03\left(1-\omega_{s}\right)^{0.173}-0.878\right]\right\}\left(1-\omega_{s}\right)^{-1} \; , \\ &
\omega_{s} \approx 1.35\; \mu_{30}^{6 / 7} M_{1}^{-2/7} \left(P L_{37}^{3 / 7}\right)^{-1} \; ; 
\end{split}
\end{align}
\cite{iaria_15} showed that $\omega_s$  is between 0.063 and 0.083 for a NS mass of X 1822-371 between 1 and 3 M$_{\odot}$.
Assuming a magnetic field strength  of  $B=(7.9\pm0.5) \times 10^{10}$ G, a NS mass of 1.61 M$_{\odot}$, a NS radius of 10 km and a spin-period derivative  
$\dot{P}_{\rm spin}$ of $-2.64(2) \times 10^{-12}$ s s$^{-1}$
we infer an intrinsic luminosity of  $1.5 \times 10^{38}$ erg 
s$^{-1}$ (corresponding to a mass accretion rate of $1.13 \times 10^{-8}$ M$_{\odot}$ yr$^{-1}$) that is very close to the Eddington luminosity of $\sim 2 \times 10^{38}$ erg 
s$^{-1}$. The intrinsic luminosity  is {20 times larger} than the observed luminosity  ({$6.6 \times 10^{36}$} erg s$^{-1}$). This discrepancy  is likely due to the high inclination angle of the source, between 81$^{\circ}$ and 84$^{\circ}$ \citep[see][]{Iaria_13, Anitra_21} which causes the emission from the central part of the system to be obscured by the outer edge of the disk. 
{Although the discrepancy between the intrinsic luminosity of the source and the observed luminosity has decreased for a source distance of 6.1 kpc, the scenario described above still holds true. As the ratio between intrinsic and observed luminosity is approximately 20, the optical depth of the ADC should be around 0.05. Therefore, the ADC remains optically thin, and all the previous conclusions remain unchanged.}

The flux associated with the blackbody component  corrected by a factor $\sim 20$ gives a blackbody radius of $190 ^{+40}_{-20}$ km {and a seed-photon radius of $10.7 \pm 0.9$ km compatible with an emission coming from the NS}. We hence associate the blackbody component to the emission from the accretion disk, its   radius indicating the inner radius of the  disk.  We do not correct this value for the inclination angle since we assume that the optically thin corona  covers the whole accretion disk.  
We infer the magnetospheric radius $R_m$ using the formula in \cite{Sanna_2017},
\begin{equation}
\label{eq:Alfv}
    R_{\mathrm{m}}=\phi\;R_{\Lambda}=\phi\;(2 G M)^{-1 / 7} \mu^{4 / 7} \dot{M}^{-2 / 7}\; ,
\end{equation}
where $G$ is the gravitational constant, $M$ is the NS mass, $\mu$ the NS magnetic dipole moment, and $\dot{M}$ is the mass accretion rate. The factor $\phi$ is given by the following equation 
\begin{equation}
    \phi=0.315 \;\kappa_{0.615}^{8 / 27}\; \alpha^{4 / 15}\; \mu_{26}^{4 / 189}\; \dot{M}_{-9}^{32 / 945}\; m^{76 / 189}, 
\end{equation}
 \citep{Sanna_2017}, 
where $\mu_{26}$, $R_{6}$ and $\dot{M}_{-9}$  are the NS magnetic moment in units of 10$^{26}$ G  cm$^{3}$, the NS  radius in units of 10$^{6}$ cm and  the mass accretion rate in unit of 10$^{-9}$ M$_{\odot}$ yr$^{-1}$. The mean molecular weight $\kappa$  is  0.615 for fully ionized matter, while the parameter $\alpha$ for a standard Shakura–Sunyaev disk model is set to 0.1 \citep{1973Shakura_Sunyaev}.  Adopting a NS mass of 1.61 M$_{\odot}$, a B-field strength of $(7.9\pm0.5) \times 10^{10}$ G and a NS radius of 10 km,  we obtain that $R_{\mathrm{m}}= 105 \pm 5$ km, that is smaller than the inner radius of the accretion disk, as expected for an X-ray pulsar. 

We observe a reflection component from the accretion disk with a ionization parameter of log$(\xi)=1.4^{+0.4}_{-1.4}$, compatible with the value 
obtained by \cite{Anitra_21}. This value  implies a relatively low ionization state of matter in the disk; the iron line is dominated by the Fe I-XX K$\alpha$ transition \citep{Fabian_2000}. The inner radius at which the reflection component is produced is $62^{+21}_{-13}  $ gravitational radii, that corresponds to $150^{+50}_{-30}  $ km  for a NS mass of 1.61 M$_{\odot}$. This value is compatible with the inner radius of the accretion disk of  $190 ^{+40}_{-20}$ km derived above. 

 The Comptonized component is associated with a optically-thick corona. 
By using eq. 10 in \cite{farinelli_08} we find that the optical depth $\tau$ is 24 and  17 for the  NICER  and \textit{NuSTAR} spectrum  assuming a spherical geometry and 12 and 8  assuming a 
slab geometry. 

From these findings, we can conclude that although there is a magnetosphere in place and the matter being accreted is channeled along the magnetic field lines, a portion of the matter has sufficient kinetic energy to overcome the magnetic barrier imposed by the magnetosphere. As a result, this matter penetrates the magnetosphere (possibly due to Rayleigh-Taylor and Kelvin-Helmholtz instabilities) and falls directly onto the surface of the NS. This phenomenon is likely favored by the large accretion rate, reaching the Eddington limit, and the magnetic field being relatively weak with a strength of $7.9 \times 10^{10}$ G.
 
We therefore suggest that the Comptonized component is formed  at  the accretion column onto the magnetic caps and the seed-photons come from all  the NS surface,  in agreement with the observation that 31\% of the seed-photons do not interact with the Comptonized region. 

 If we assume that the  broadening of the cyclotron line 
 ($\sigma_0= 0.17\pm0.03$ keV) has  thermal origin, we estimate a temperature of 13 keV   not consistent with the electron temperature of 4.3  keV obtained from the fit.  A possible explanation is that we are observing a gradient of the $B$-field strength along the accreting column. In this hypothesis we can infer the height of the column considering that the magnetic dipole moment   $\mu=B R^3$ does not change, and then $-\Delta B/B = 3\Delta R/R$. By imposing  $\Delta B$ as the sigma of the  {\tt gabs} component we find that  $ \Delta R \simeq 0.9$ km, assuming a NS radius of 10 km. 

  Assuming a fully ionized hydrogen plasma ($n_e=n_p$) in the accretion column, the proton density 
 is given by $n_p \sim 10^{21} \dot{M}_{18}/(\beta S_{10})$ cm$^{-3}$ \citep[see eq. 1 in][]{Mushtukov_19}, where $\dot{M}_{18}$ is the mass accretion rate in units of $10^{18}$
 g s$^{-1}$, $S_{10}$ is the cross section of the accretion column in units of  $10^{10}$ cm$^{2}$ and $\beta=v/c$
with $v$   the local velocity of plasma  and $c$ the speed of the light.  Since
the accreting matter   slows down from a free-fall velocity $v_{ff}$  to about $0.1 v_{ff}$  in the radiation dominated shock at the top of accretion column we assume $v=0.1 v_{ff}$ , that is $\beta \simeq 0.07$. The cross section of the accretion column, i. e. the area on the star over which the accretion occurs, is $S\simeq R^3 R_m^{-1}$ cm$^{2}$ \citep{Pringle_72}, where $R$ is the NS radius. For a NS radius of 10 km 
we obtain $S\simeq 8.3 \times 10^{10}$ cm$^{2}$. Using the mass accretion rate estimated from eq. 3 we find that $n_p \simeq 1.2 \times 10^{21}$ cm$^{-3}$. Since $n_e=n_p$ we can estimate the thickness $l$ of the Comptonized region in the accretion column using 
$\tau=\sigma_T n_e l$  where $\sigma_T$ is the Thomson cross section. In the reasonable assumption that the accretion column has a slab geometry we can adopt an optical depth  $\tau =8$ estimated above. We find $l \simeq 0.1$ km, that is the Comptonizing region   has a thickness that is 1/8 of the height of the accretion column.

Finally, \cite{jonker_01} observed that the pulse fraction increases going from 5 keV to 20 keV,  this can be explained
in our scenario considering that the spin modulation is associated with the accretion column from which upscattered photons escape. The observed pulsation is due to the Comptonized photons leaving the accretion column which are modulated by the NS rotation. In this scenario, we do not observe a spin modulation at   energies below 5 keV because this range is dominated by the soft thermal components corresponding to the inner disk and the seed-photons that come from all over the NS surface.

Our model suggests that the Comptonization component primarily originates from the accretion column. From our best-fit to the data, we find that the reflection amplitude is $f_{refl} = 0.66 ^{+0.08}_{-0.10}$ \citep[a comparable  value of $0.61^{+0.04}_{-0.05}$ was obtained by][]{Anitra_21}.
Typical values of the reflection amplitude $\Omega/2\pi$ for NS-LMXB atoll sources range from 0.2-0.3 \citep[see, e.g.,][]{Disalvo_15,Disalvo_19,Dai_13,Egron_13,Matranga_17,Marino_19} and suggest the presence of a spherical corona in the inner part of the accretion disk \citep[see Fig. 5 in][]{Dove_97}. The large reflection amplitude could be due to the peculiar geometry, in our scenario the radiation incident the disk (truncated at the magnetospheric radius) comes from a compact region located at the surface of the neutron star. 

\cite{Iaria_13} proposed that the \ion{Fe}{i} and the  \ion{Fe}{xxvi} emission lines are  produced in the photoionized surface of the accretion disk at a distance from the NS of $2 \times 10^{10}$ cm and $< 3.7 \times 10^{10}$ cm, respectively. 
However, the ionization parameter $\xi$ in the inner region of the accretion disk, where the reflection component originates, is close to 25, which is too low to produce \ion{Fe}{xxvi} ions.
We suggest that these narrow lines are produced at large distance from the central source, possibly in a disk atmosphere with a large gradient in temperature and/or photoionization, although further investigations are necessary to address their origin. 

The absorption edge at $9.62 \pm 0.09$ keV is not compatible 
with   a \ion{Fe}{xxvi} transition expected at 9.278 keV in a rest frame.  If we interpret this feature as  blue-shifted  we derive a wind speed of 11,000 km/s. However,   \cite{Anitra_21}  suggested that this feature could have a systematic origin.

 Finally, we observe the presence of a power-law component never observed before in this source. This component looks similar to the power-law hard tail often observed in the X-ray spectra of bright LMXBs, both in Z-sources \citep[see, e.g.,][]{disalvo_00,Iaria_01_cir,disalvo_01} and in atoll sources in their soft states   \citep[see, e.g.,][and references therein]{piraino_07}. When significantly detected, these components usually show a power-law shape with a photon index $\sim 2-3$ contributing to a few percent of the total flux from the source, and sometimes a correlation with the radio emission has been observed \citep[e.g.,][]{Homan_04,MIgliari_07}.
  The origin of these features is still debated, but it is probably related to the presence of electrons with a non-thermal velocity distribution \citep[e.g.,][]{disalvo_06}. Considering that X 1822-371 is an X-ray pulsar, we tentatively suggest that non-thermal electrons moving along the magnetic field lines towards the NS polar caps may be responsible for the observed power-law hard tail in this source.

\section{Conclusions}
We  analyzed  a 0.3-50 keV broadband spectrum of X 1822-371 combining a   NICER  spectrum (0.3-10 keV) with a  \textit{NuSTAR} spectrum (4.5-50 keV).   Our best-fit to the spectrum gives  clear evidence of an absorption cyclotron line 
 with energy  close to $0.66$ keV, 
confirming the detection reported by \cite{iaria_15}. From 
the temporal analysis we find that:
\begin{itemize}
    \item The orbital period is expanding at a constant derivative of
    $\dot{P}_{\rm orb}= 1.426(26) \times 10^{-10}$ s s$^{-1}$,
    \item the NS is spinning up with a long-term spin-period derivative of $\Dot{P}_{spin} = -2.64(2) \times 10^{-12}$ s s$^{-1}$ (with  ${P}_{spin}=0.59112(2)$  s),
    \item the observed absorption cyclotron line  close to 0.66 keV  corresponds to a magnetic field strength of 
    $B=(7.9\pm0.5) \times 10^{10}$ G  for a NS mass of 1.61 M$_{\odot}$,  
    \item using the $\Dot{P}_{spin}$ and $B$ values we find that the luminosity of the source is $1.5 \times 10^{38}$ erg s$^{-1}$ and the magnetospheric radius   $R_{\mathrm{m}}= 105 \pm 5$ km, 
    \item because the observed  0.1-100 keV  luminosity is a factor   20 smaller than  $1.5 \times 10^{38}$ erg s$^{-1}$ and the inclination angle of the system is between 81$^{\circ}$ and 84$^{\circ}$,  we suggest the presence of an extended  optically thin corona (with optical depth of {0.05}) that scatters along the line of sight about {5\%} of the radiation produced in the inner region of the system.
\end{itemize}
 By correcting the observed flux by a factor {20} we associate the thermal component with a thermal emission from the accretion disk finding an inner radius of  $190^{+40}_{-20}$ km. The reflection component is produced at an inner radius of $150^{+50}_{-30}  $ km, that is compatible with the inner radius of the accretion disk.  

The Comptonized  region is associated with the accretion column at the magnetic caps. We find that the equivalent spherical radius of the seed-photons emission region is $10.7 \pm 0.9$
km, that is very similar to the NS radius;  we find that  31\% of these soft photons do not interact with the Comptonizing region. 

The broadening of the cyclotron line can be explained with a gradient of the magnetic field $B$ associated with  a height of the accretion column of $\sim 0.9$ km. We estimate an average electron density of  $1.2 \times 10^{21}$ cm$^{-3}$ in the accretion column.  

We observe a reflection component with an  amplitude  of $0.66 ^{+0.08}_{-0.10}$, about a factor of $\sim 2$ larger than the typical values of $0.2-0.3$ observed in the NS-LMXBs. A possible explanation is that    the  incident radiation to the disk comes from  the  photons  leaving the accretion column with a geometry roughly similar to the lamp-post geometry observed in AGNs.  

Finally, we suggest that the two observed narrow lines associated with the presence of neutral iron and \ion{Fe}{xxvi} ions  form in a
region of the system, possibly a disk atmosphere, likely at a large distance,  $\sim 2 \times 10^{10}$ cm, from the central source.

\begin{acknowledgements}The authors acknowledge financial contribution from the agreement ASI-INAF n.2017-14-H.0 and INAF mainstream (PI: A. De Rosa, T. Belloni), from the HERMES project financed by the Italian Space Agency (ASI) Agreement n. 2016/13 U.O and from the ASI-INAF Accordo Attuativo HERMES Technologic Pathfinder n. 2018-10-H.1-2020. We also acknowledge support from the European Union Horizon 2020 Research and Innovation Frame- work Programme under grant agreement HERMES-Scientific Pathfinder n. 821896 and from PRIN-INAF 2019 with the project "Probing the geometry of accretion: from theory to observations" (PI: Belloni).

\end{acknowledgements}
\bibliographystyle{aa}
\bibliography{biblio}
\end{document}